\documentclass[
  manuscript=article,  
  layout=preprint,  
  year=20xx,
  volume=x,
  12pt, footnote=numbered,
]{extra/joas}

\conference{Conference Title}

\received {1 April 20xx}
\revised  {1 May 20xx}
\accepted {10 May 20xx}
\published{20 May 20xx}

\editor{Editor Name}

\reviewers{First Reviewer, Second Reviewer, Third Reviewer}

\usepackage[english]{babel} 
\usepackage{blindtext}
\usepackage{multirow}
\usepackage{longtable}
\usepackage{array}
\usepackage{colortbl}
\usepackage{xcolor}
\usepackage{makecell}
\usepackage{lscape}
\usepackage[backend=biber]{biblatex}

\usepackage{amssymb}

\makeatletter
\patchcmd{\@maketitle}
  {\hfill\includegraphics[width=20mm]{}\par}
  {\hfill\par}
  {}{}
\makeatother

\usepackage{titlesec}

\titleformat{\paragraph}
  {\normalfont\normalsize}  
  {\theparagraph}  
  {1em}  
  {}     

\titlespacing{\paragraph}
  {0pt}  
  {3.25ex plus 1ex minus .2ex}  
  {1.5ex plus .2ex}  
\RequirePackage{pdflscape}
\RequirePackage{fancyhdr}
\RequirePackage[absolute]{textpos}
\fancypagestyle{lscape}{
    \fancyhf{}
    \fancyhead[L]{
        \rotatebox{90}{\normalsize \thepage}
    }%
    \addtolength{\headheight}{2cm}
    \addtolength{\headsep}{-1.5cm}
}

\bibliography{reference}

\title{A Review of LLM-Assisted Ideation}

\author{Sitong Li \orcid{0009-0001-4985-8097}}
\affiliation{Heriot-Watt University, Edinburgh, UK}

\author{Stefano Padilla \orcid{0000-0002-7104-8349}}
\affiliation{Heriot-Watt University, Edinburgh, UK}

\author{Pierre Le Bras \orcid{0000-0002-0670-6552}}
\affiliation{Heriot-Watt University, Edinburgh, UK}

\author{Junyu Dong \orcid{0000-0001-7012-2087}}
\affiliation{Ocean University of China, Qingdao, China}

\author{Mike Chantler \orcid{0000-0002-8381-1751}}
\affiliation{Heriot-Watt University, Edinburgh, UK}
\email{M.J.Chantler@hw.ac.uk}

\keywords{large language models; LLMs; ideation; creativity support; interactive systems; interaction design; user study.}

\begin{document}
\begin{abstract}
We present a comprehensive, in-depth review of ideation assisted by large language models (LLMs), highlighting emerging trends and identifying unaddressed research gaps. In total, we examined 61 studies investigating the application of LLMs in both group and individual ideation processes. From these studies, we derived the Hourglass Ideation Framework  for LLM-assisted ideation, comprising three phases and seven key ideation stages, which served as the basis for our systematic survey of relevant publications. Our analysis reveals that LLMs are most frequently used for idea generation and refinement, but their use in scope specification, foundational material structuring and multi-idea evaluation and selection remains limited. We provide our findings in extensive tabular and online formats \footnote[1]{Resources are available at \url{https://doi.org/10.6084/m9.figshare.28440182}}. These catalogues detail research on LLM-assisted, purely LLM-based, and human-only activities across the seven ideation stages for each of the 61 studies. These also detail creative domains, publication outlets, interaction designs, user study designs, and assessment methods. Our analysis of system interaction design reveals a predominant focus on supporting individual ideation activities and text-based interaction, with a growing trend of incorporating multimedia elements. However, in group ideation, tools and interaction modalities targeting both synchronous and asynchronous collaboration are much scarcer. We synthesize the primary findings of our review and outline promising directions for future research in LLM-assisted ideation. We hope this review will help researchers quickly gain an overview of this rapidly expanding area, efficiently locate relevant work, and rapidly identify underexplored areas for further investigation. In addition, we believe that the framework we present here will form the basis for the development of future problem and solution space taxonomies, and associated methodologies for LLM-assisted ideation development and use.
\end{abstract}

\section{Introduction}

Ideation, the process of generating and developing ideas, is a critical driver of innovation and creativity across diverse domains \cite{osborn1953applied, amabile1988}. Over the years, many techniques and tools have been developed to facilitate ideation, ranging from traditional methods such as brainstorming and the Delphi method to assisted ideation approaches, including machine-driven analytical techniques and idea management systems. However, these approaches often face challenges in generating a wealth of truly novel and valuable ideas, overcoming cognitive biases and fixation effects, and supporting efficient evaluation and selection of potentially many promising ideas \cite{smith2003, blair2007, acar2019}. The emergence of large language models (LLMs) has brought new possibilities for ideation. LLM-assisted ideation has demonstrated its advantages in multiple domains, with LLMs not only being capable of assisting ideation in all stages, but also being widely integrated into systems to fully exploit their potential. Nevertheless, there is currently a lack of research systematically analyzing the stages at which LLMs can assist ideation and how they may do so. Moreover, no studies have yet comprehensively examined how LLMs are integrated into systems to interact with humans. Investigating these aspects can help deepen our understanding of the paradigms of LLM-assisted ideation, facilitate the better use of LLMs to support ideation, and contribute to the development of more efficient ideation systems.

\subsection{Background}
\subsubsection{Traditional Ideation}

One of the earliest and most well-known ideation techniques is brainstorming, introduced by Alex Osborn in the 1950s \cite{osborn1953applied}. Brainstorming aims to stimulate creative thinking by encouraging the free generation of a large quantity of ideas in a non-judgmental environment. Since its introduction, brainstorming has been widely adopted and has inspired numerous variations and extensions, such as electronic brainstorming \cite{dennis1993} and brainwriting \cite{heslin2009}. A recent literature review by \cite{maaravi2021ideation} analysed the advantages and disadvantages of electronic brainstorming and proposed a comprehensive model to support creative ideation.

Other classic ideation techniques include the Delphi method \cite{dalkey1963}, which seeks to achieve consensus among a panel of experts through iterative rounds of anonymous idea generation and feedback, and morphological analysis \cite{zwicky1969}, which systematically explores the combinatorial space of possible solutions by decomposing a problem into its constituent dimensions. These techniques, along with many others, have been extensively studied and applied in various domains to support ideation \cite{vangundy1988}.

However, traditional ideation techniques often face challenges in efficiently navigating large problem spaces, overcoming cognitive biases and fixation effects, generating a wide range of novel ideas, and providing structured support for elaborating and refining ideas \cite{kohn2011collaborative, brem2007innovation}.

\subsubsection{Assisted Ideation}

In recent years, the rapid advancement of digital technologies has opened up new frontiers for assisted ideation. One notable development is the emergence of machine-driven analytical techniques, such as machine learning, to support idea generation. A comprehensive literature review by \cite{ayele2021systematic} identified and summarized over 20 data-driven and machine-driven techniques for enhancing idea generation, highlighting the potential of these approaches to augment human creativity.

Another significant trend is the development and adoption of idea management systems, which provide digital platforms for capturing, organizing, and managing ideas. A recent literature review by \cite{zhu2023taxonomy} classified and summarized over 30 current idea management tools, underscoring their growing prominence. Idea management systems offer numerous benefits, such as electronically storing ideas to prevent loss and improve accessibility \cite{bakker2006, boeddrich2004, brem2009}, enabling ideators to search for and build upon existing ideas \cite{sandstrom2010, stenmark2000, wilson2010, xie2010}, and facilitating the submission and management of ideas through standardized templates and categorization schemes \cite{fairbank2003, bothos2008, gamlin2007, vahs2013, xie2010}.

Despite these advancements, assisted ideation still faces significant challenges. Machine learning techniques, such as using deep learning to analyse patent documents for generating design concepts \cite{hope2017accelerating}, are powerful but typically require large training datasets and may struggle to generate truly novel and creative ideas \cite{garfield2001}. While idea management systems have greatly improved the capture and organization of ideas, they often lack advanced capabilities for stimulating idea generation, expanding and refining ideas, and supporting the evaluation and selection of promising ideas \cite{acar2019}.

\subsubsection{LLM-assisted Ideation}

In recent years, the emergence of large language models (LLMs) has opened up exciting new possibilities for assisted ideation. LLMs, such as GPT-3 \cite{brown2020}, are AI systems trained on vast amounts of text data, enabling them to generate human-like text, answer questions, and perform a wide range of language tasks. The potential of LLMs to enhance ideation is immense, as they can serve as powerful tools for generating novel ideas by breaking free from habitual patterns of thought, overcoming fixation effects, and navigating vast problem spaces \cite{smith2003}. Additionally, LLMs can aid in expanding and refining concepts, and even assist in the evaluation and selection of potentially numerous promising ideas by mitigating cognitive biases \cite{maaloe2022, wang2022}.

\subsection{Related Work}

Study \cite{gerlach2017idea} proposed a conceptual framework for idea management that provides a structured approach to systematically capture, develop, and implement innovative ideas within organizations, forming a generic model for enterprise idea management projects. This conceptual framework consists of six main phases: preparation, idea generation, improvement, evaluation, implementation, and deployment. However, their framework only considers the enterprise idea management level, neglecting other groups such as individuals and organizations, and does not consider the conceptual framework at a more general ideation level. Importantly, it was developed before the advent of LLMs and does not consider their unique capabilities and challenges. With the rapidly growing interest in LLMs for ideation, there is currently a lack of frameworks that identify the different stages involved in such processes and guide the development and application of LLM-assisted ideation.

Study \cite{zhu2023taxonomy} classified and summarized current idea management tools and highlighted the importance of studying user interaction and experience with idea management tools. However, their review did not consider the recent emergence of large language models (LLMs) and their potential applications in ideation processes. It is clear that a thorough analysis of how users interact with LLM-assisted ideation systems could provide valuable insights to inform the development of more effective and user-friendly systems. Given the novelty of LLMs and their potential to facilitate ideation, understanding user interaction patterns, preferences, and challenges in the context of LLM-assisted ideation is crucial.

\subsection{Contributions}

To address the above gaps, this review aims to provide an analysis of the current state of research on LLM-assisted ideation and propose a framework to help facilitate future research and development in this field. Then, we make three main contributions:

\begin{itemize}
    \item [1.] We provide a comprehensive overview of the current state of LLM-assisted ideation research containing publication statistics, application domains, and ideation methods. The findings provide a foundation for understanding the development of the field.
    
    \item [2.] We propose a framework for LLM-assisted ideation derived from our literature review, that identified seven stages of the ideation process. We use this to understand how LLMs have been used to support ideation.
    
    \item [3.] We analyse the interaction design and user study of LLM-assisted ideation systems based on the relevant literature. The insights gained from this analysis can serve as valuable guidance for designing and developing more user-friendly and efficient LLM-assisted ideation systems.

\end{itemize}

\section{Methodology}

\subsection{Inclusion and Exclusion Criteria}

This literature review aims to investigate the recent developments in the application of large language models (LLMs) to support ideation. Ideation, the process of generating and articulating novel ideas, is a critical phase in the innovation process and serves as the foundation for groundbreaking advancements. LLMs, with their remarkable natural language processing capabilities, have emerged as a powerful tool in artificial intelligence, offering the potential to enhance and accelerate human ideation efforts. Consequently, examining the intersection and synergies between these two key areas forms the primary focus of this review.

The inclusion criteria encompassed scholarly works that investigated the use of LLMs to support ideation processes in diverse contexts or tasks, such as design, writing, and multimedia creation. In contrast, the exclusion criteria consisted of: (1) books, workshops, review articles, or brief overview papers; (2) studies that addressed ideation-related contexts or tasks but did not employ LLM methods; (3) research that utilized LLM techniques but did not focus on ideation stages.

\begin{figure}[ht!]
  \centering
  \includegraphics[width=0.7\textwidth]{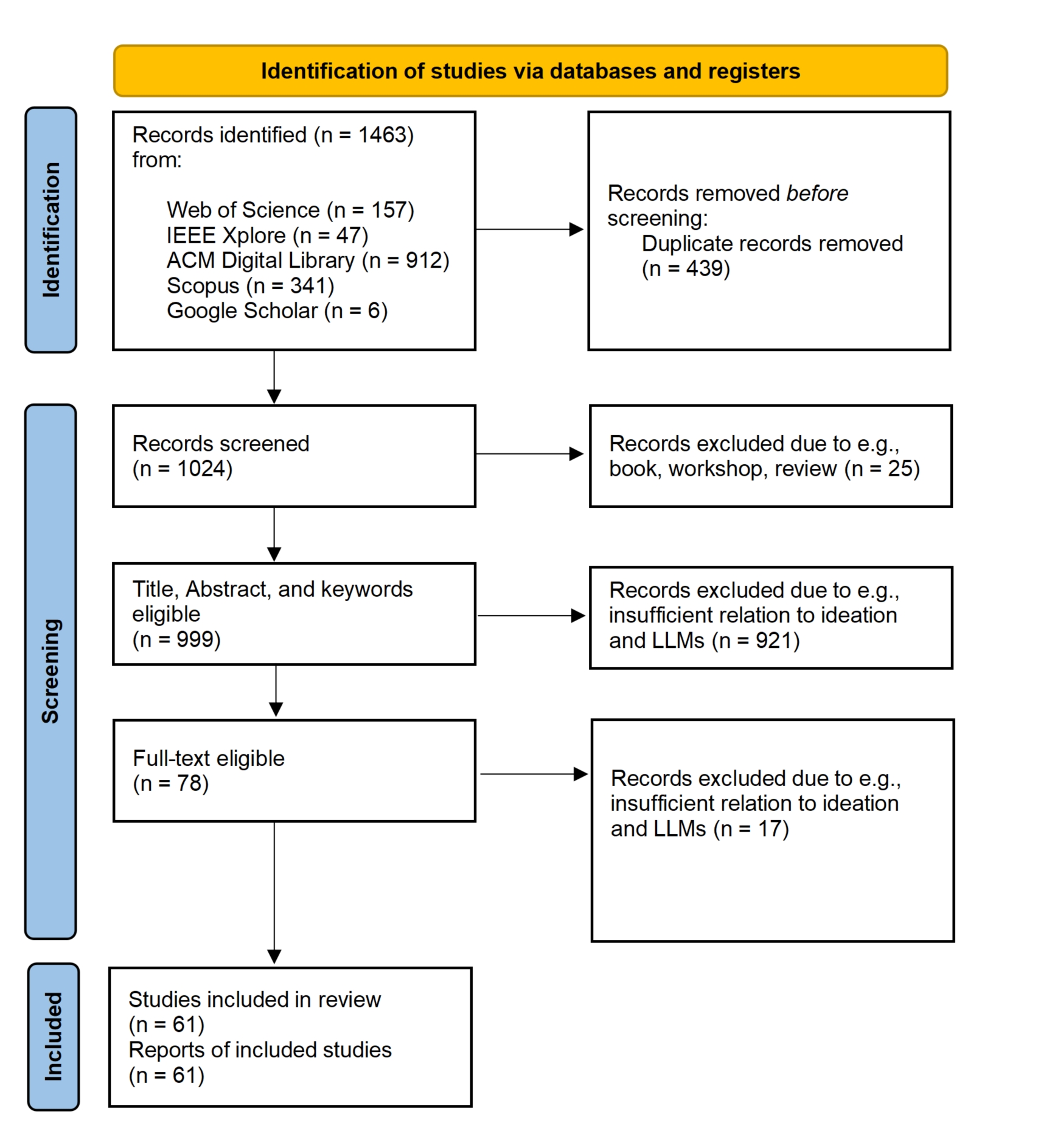}
  \caption{PRISMA Flow Diagram for Screening}
  \label{fig:Screening process}
\end{figure}

\subsection{Search Strategy}

To ensure a comprehensive and representative literature search, four prominent databases were consulted: Web of Science, IEEE Xplore, ACM Digital Library, and Scopus. These databases cover mainstream journals and conference proceedings within the relevant fields of computer science, artificial intelligence, human-computer interaction, and related disciplines, aligning with the thematic scope of this study. The literature search primarily employed an abstract search approach, using keywords such as "ideation," "idea generation," "large language models," and their synonymous variants.

\subsection{Screening}

To maintain systematic rigor and adherence to established protocols, the evaluation process followed the PRISMA criteria for systematic reviews \cite{moher2009preferred}, adapted to incorporate the recommendations proposed by Page et al. \cite{page2021prisma}. The screening process is comprised of the following steps, as depicted in Figure \ref{fig:Screening process}. Initially, a comprehensive search across four databases identified 1457 papers, supplemented by an additional 6 papers from Google Scholar, resulting in a total of 1463 records. After removing 439 duplicates, 1024 unique documents were retained for subsequent screening. The first round of screening eliminated 25 books, workshops, or reviews, retaining 999 papers. The second round involved examining titles, abstracts, and keywords to exclude 921 papers unrelated to "ideation" and "large language models," preserving 78 papers. The third round consisted of full-text screening, further excluding 17 papers that did not meet the inclusion criteria, ultimately retaining 61 papers for analysis.

\subsection{Data Extraction and Analysis Method}

The data extraction and analysis processes employed in this literature review involved a manual approach to obtain key information from the selected articles. A comprehensive text analysis was conducted on the full text of each included study. Essential elements were systematically collected from every article, such as publication details, application domains, ideation methods, ideation stages, interaction forms, and user study methods. The extracted data were then synthesized to derive insights from multiple analytical perspectives, aligning with the primary themes identified in the literature review. 

By adhering to this rigorous methodology, this literature review aims to provide a comprehensive and systematic analysis of the current state of research on LLM-assisted ideation, identifying key trends, challenges, and opportunities for future work in this rapidly evolving field. The insights gained from this review will contribute to a deeper understanding of how LLMs can be effectively leveraged to support and enhance ideation creativity and innovation across various domains.

\section{Overview}
\subsection{Publication Statistics}

\begin{figure}[ht!]
  \centering
  \includegraphics[width=1\textwidth]{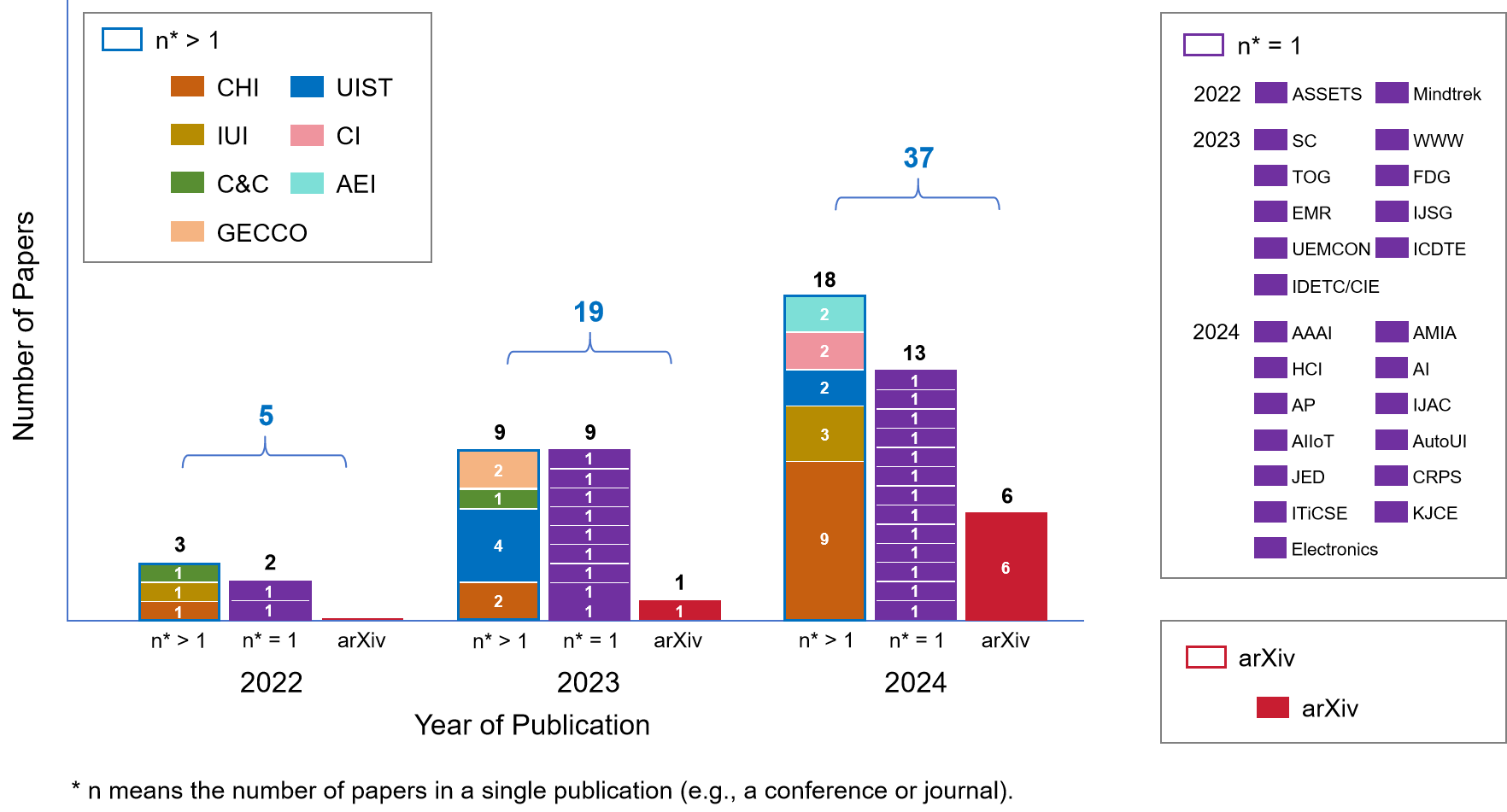}
  \caption{Publication Statistics of Analysed Papers}
  \label{publication}
\end{figure}

This review considered 61 scholarly works, sourced from 31 journals and conference proceedings, as well as the arXiv preprint repository. As shown in Figure \ref{publication},  these venues span a range of disciplines, including human-computer interaction, computational creativity, engineering, and artificial intelligence. Among the 31 publishing venues, 7 have published 2 or more articles related to LLM-assisted ideation. The \textit{ACM Conference on Human Factors in Computing Systems (CHI)} is the most prominent outlet, with 12 articles, followed by the \textit{ACM Symposium on User Interface Software and Technology (UIST)} with 6 articles, and the \textit{ACM International Conference on Intelligent User Interfaces (IUI)} with 4 articles. The \textit{ACM Collective Intelligence Conference (CI)}, \textit{ACM Conference on Creativity \& Cognition (C\&C)}, \textit{Advanced Engineering Informatics (AEI)} and \textit{Conference on Genetic and Evolutionary Computation (GECCO)} each published 2 articles. Additionally, 24 other venues contributed a single article each, and 7 preprints were included from arXiv.

Notably, all articles were published within the last three years. The number of published articles has increased significantly over time, from just 5 articles in 2022 to 19 articles in 2023, and reaching 37 articles in 2024, reflecting the rapidly growing interest in this research area. Moreover, the number of relevant preprints on arXiv has also been increasing over the past two years. These trends suggest that the field of LLM-assisted ideation holds immense potential and is currently in a phase of rapid development, attracting researchers from a variety of disciplines to contribute to its advancement.

\subsection{Domains and Application Areas}

The literature on LLM-assisted ideation covers a wide range of domains and application areas, demonstrating the versatility and potential of this technology. As shown in Table \ref{tb:Domains and Application Areas}, the reviewed articles can be broadly categorized into five main domains: design, writing, multimedia creation, scientific research, and education.

In the design domain, LLMs have been applied to various sub-domains, including general design, product design, game design, visual design, sustainable design, and biological design. These applications encompass a diverse set of downstream tasks, such as project design \cite{kocaballi2023conversational_D1}, prototyping \cite{bilgram2023accelerating_D2,ege2024chatgpt_D10}, game level design \cite{sudhakaran2023prompt_D3,todd2023level_D4}, interior color design \cite{hou2024c2ideas_D13}, eco-design \cite{lee2024generating_D7}, and protein design \cite{dharuman2023protein_D5}.

The writing domain showcases the use of LLMs in general writing tasks \cite{kim2024towards_W2,kim2023cells_W9,reza2024abscribe_W13,benharrak2024writer_W14,yuan2022wordcraft_W5}, as well as more specific applications such as prewriting \cite{wan2024felt_W8,petridis2023anglekindling_W6}, creative and narrative writing \cite{chung2022talebrush_W3,ghajargar2022redhead_W7,li2024mystery_W11}, academic and professional writing \cite{pividori2024publishing_W12,zhang2023visar_W4,goldi2024intelligent_W10}, and digital and informal writing \cite{goodman2022lampost_W1,kim2023cells_W9,reza2024abscribe_W13}.

In the multimedia creation domain, LLMs have been employed for visual content generation, including image generation \cite{cai2023designaid_M1,wang2023popblends_M2,brade2023promptify_M3,chen2024foundation_M8,kang2024biospark_G14,huang2024plantography_D14}, picture book generation \cite{wang2023script_M6}, shape generation \cite{qian2024shape_M5}, and icon generation \cite{wu2023iconshop_M7}. Additionally, LLMs have been used for video editing \cite{wang2024lave_M4}.

The scientific research domain highlights the application of LLMs in generating research ideas \cite{pu2024ideasynth_S1,liu2024personaflow_S2,radensky2024scideator_S3}, as well as in more specific fields such as social psychology \cite{banker2024machine_S4}, materials science \cite{chen2024use_S5}, and chemical process design \cite{lee2024gpt_S6}.

Lastly, the education domain showcases the use of LLMs in instructional design and tools \cite{junior2023chatgpt_D11,huang2023causalmapper_G5,veloso2024forming_D12}, academic writing support \cite{goldi2024intelligent_W10}, knowledge building \cite{lee2024prompt_G18}, and understanding computing concepts \cite{bernstein2024like_G11}.

\newpage
\begin{longtable}{>{\raggedright\arraybackslash}p{1.5cm} >
  {\raggedright\arraybackslash}p{4cm} >
  {\raggedright\arraybackslash}p{4.5cm} >
  {\raggedright\arraybackslash}p{2.5cm} }
  \caption{Domains and Application Areas of LLM-assisted Ideation} \label{tb:Domains and Application Areas} \\ 
  \toprule
  \textbf{Domain} & \textbf{Application Area} & \textbf{Downstream Task} & \textbf{Ref.} \\
  \midrule
  \endfirsthead
  \\
  \\
  \toprule
  \textbf{Domain} & \textbf{Application Area} & \textbf{Downstream Task} & \textbf{Ref.} \\
  \midrule
  \endhead
  \multirow{12}{*}{\textbf{Design}} & \multirow{2}{*}{\textbf{General Design}} & project design & \cite{kocaballi2023conversational_D1} \\
  & &  conceptual design & \cite{wang2023task_D8} \\ 
  \cmidrule{2-4}
  & \multirow{2}{*}{\textbf{Product design}} 
  & prototyping design & \cite{bilgram2023accelerating_D2}, \cite{ege2024chatgpt_D10} \\
  & & interactive devices design & \cite{lu2024large_D9} \\
  \cmidrule{2-4}
  & \multirow{3}{*}{\textbf{Game Design}} & game design & \cite{lanzi2023chatgpt_D6} \\
  & & game levels design & \cite{sudhakaran2023prompt_D3}, \cite{todd2023level_D4} \\ 
  & & educational board games design & \cite{junior2023chatgpt_D11} \\  
  \cmidrule{2-4}
  & \multirow{3}{*}{\textbf{Visual Design}} & interior color design & \cite{hou2024c2ideas_D13} \\
  & & building envelope design & \cite{veloso2024forming_D12} \\
  & & landscape renderings design & \cite{huang2024plantography_D14} \\
  \cmidrule{2-4}
  & \textbf{Sustainable Design} & eco-design & \cite{lee2024generating_D7} \\
  \cmidrule{2-4}
  & \textbf{Biological Design} & protein design & \cite{dharuman2023protein_D5} \\

  \cmidrule{1-4}
  \multirow{17}{*}{\textbf{Writing}} & \textbf{General Writing} & general writing & \cite{kim2024towards_W2}, \cite{kim2023cells_W9}, \cite{reza2024abscribe_W13},  \cite{benharrak2024writer_W14},  \cite{yuan2022wordcraft_W5} \\
  \cmidrule{2-4}   
  & \multirow{2}{*}{\textbf{Prewriting}} & general prewriting & \cite{wan2024felt_W8}\\
  & & develop press release angles & \cite{petridis2023anglekindling_W6} \\
  \cmidrule{2-4}  
  & \multirow{3}{*}{\textbf{Creative and Narrative Writing}} & story writing & \cite{chung2022talebrush_W3}, \cite{yuan2022wordcraft_W5}, \cite{wan2024felt_W8}, \cite{kim2023cells_W9} \\
  & & fiction writing & \cite{ghajargar2022redhead_W7}, \cite{benharrak2024writer_W14} \\
  & & mystery game script writing & \cite{li2024mystery_W11} \\
  \cmidrule{2-4}  
  & \multirow{5}{*}{\textbf{Academic and Professional Writing}} & scholarly writing & \cite{pividori2024publishing_W12}, \cite{benharrak2024writer_W14}, \cite{yuan2024llmcrit_W15} \\
  & & argumentative writing & \cite{zhang2023visar_W4}, \cite{kim2024towards_W2} \\    
  & & student peer review writing & \cite{goldi2024intelligent_W10} \\
  & & report writing & \cite{benharrak2024writer_W14} \\
  & & proposal writing & \cite{kim2024towards_W2}, \cite{benharrak2024writer_W14} \\
  \cmidrule{2-4}   
  & \multirow{6}{*}{\textbf{Digital and Informal Writing}} & email & \cite{goodman2022lampost_W1},\cite{kim2023cells_W9},\cite{reza2024abscribe_W13},\cite{benharrak2024writer_W14} \\
  & & blog post & \cite{kim2024towards_W2}, \cite{benharrak2024writer_W14}, \cite{yuan2024llmcrit_W15} \\
  & & social media post & \cite{reza2024abscribe_W13} \\
  & & copywriting & \cite{kim2023cells_W9} \\  
  & & web article & \cite{benharrak2024writer_W14} \\
  & & newsletter & \cite{kim2024towards_W2}, \cite{petridis2023anglekindling_W6} \\
  
  \cmidrule{1-4}
  \multirow{7}{*}{\shortstack{\textbf{Multimedia} \\ \textbf{Creation \phantom{11}}}} & \multirow{6}{*}{\textbf{Visual Content Generation}} & image generation & \cite{cai2023designaid_M1}, \cite{wang2023popblends_M2}, \cite{brade2023promptify_M3}, \cite{chen2024foundation_M8}, \cite{kang2024biospark_G14}, \cite{huang2024plantography_D14} \\
  & & image color matching & \cite{hou2024c2ideas_D13} \\
  & & picture book generation & \cite{wang2023script_M6} \\
  & & shape generation & \cite{qian2024shape_M5} \\
  & & icon generation & \cite{wu2023iconshop_M7} \\
  \cmidrule{2-4}
  & \textbf{Video Editing and Creation} & video editing & \cite{wang2024lave_M4}\\

  \cmidrule{1-4}
  \multirow{7}{*}{\shortstack{\textbf{Scientific} \\ {\textbf{Research}}}} & \textbf{General Scientific Research} & research idea generation & \cite{pu2024ideasynth_S1}, \cite{liu2024personaflow_S2}, \cite{radensky2024scideator_S3} \\ 
  \cmidrule{2-4}
  & \multirow{2}{*}{\textbf{Social Sciences}} & social psychology hypothesis generation & \cite{banker2024machine_S4} \\
  \cmidrule{2-4}
  & \multirow{2}{*}{\textbf{Materials and Experimental Sciences}}  & experimental materials hypothesis generation & \cite{chen2024use_S5} \\ 
  \cmidrule{2-4}
  & \textbf{Chemical Sciences}  & chemical process design improvement & \cite{lee2024gpt_S6} \\  

  \cmidrule{1-4}
  \multirow{7}{*}{\textbf{Education}} & \multirow{3}{*}{\textbf{Instructional Design and Tools}} & educational board game design & \cite{junior2023chatgpt_D11} \\ 
  & & systems thinking education & \cite{huang2023causalmapper_G5} \\
  & & architecture design education & \cite{veloso2024forming_D12} \\ 
  \cmidrule{2-4}
  & \textbf{Academic Writing}  & writing process support & \cite{goldi2024intelligent_W10} \\
  \cmidrule{2-4}
  & \textbf{Knowledge Building}  & knowledge creation process & \cite{lee2024prompt_G18} \\
  \cmidrule{2-4}
  & \textbf{Knowledge Learning}  & computing concepts understanding & \cite{bernstein2024like_G11} \\
  \bottomrule
\end{longtable}

\subsection{General Domain and Ideation Methods}

Beyond the five specific domains discussed in the previous section, a significant body of research has explored the application of LLMs in supporting ideation activities from a more general perspective. These studies have investigated the integration of ideation methods with LLMs to guide and facilitate the process. As illustrated in Table \ref{tb:Ideation Method of General Domain}, the ideation methods employed in these studies can be broadly categorized into 7 main types: structured thinking, decomposition and composition, analogical thinking, multi-dimensional thinking, information-driven thinking, scenario-based thinking, and collaborative ideation. By leveraging these diverse ideation methods in conjunction with LLMs, researchers aim to enhance the effectiveness and efficiency of the ideation process.

\begin{longtable}{>{\raggedright\arraybackslash}p{2.2cm} >
  {\raggedright\arraybackslash}p{3.3cm} >
  {\raggedright\arraybackslash}p{4.5cm} >
  {\raggedright\arraybackslash}p{2.5cm} }
  \caption{Ideation Methods of General Domain} \label{tb:Ideation Method of General Domain} \\ 
  \toprule
  \textbf{Ideation Method} & \textbf{Specific Method} & \textbf{Specific Content} & \textbf{Ref.} \\
  \midrule
  \endfirsthead
  \\
  \\
  \toprule
  \textbf{Ideation Method} & \textbf{Specific Method} & \textbf{Specific Content} & \textbf{Ref.} \\
  \midrule
  \endhead
  \multirow{8}{*}{\shortstack{\textbf{Structured} \\ \textbf{Thinking \phantom{1}}}} & \multirow{4}{*}{\textbf{Thinking Chain Method}} & chain of thought & \cite{lee2024prompt_G18}, \cite{hou2024c2ideas_D13} \\
  & & graph of thoughts & \cite{besta2024graph_G12} \\
  & & causal map & \cite{huang2023causalmapper_G5} \\
  & & mind map & \cite{liu2024personaflow_S2}, \cite{zhang2023visar_W4} \\  
  \cmidrule{2-4}
  & \multirow{4}{*}{\textbf{Structured Template}} & template for reflection  & \cite{xu2024jamplate_G3} \\ 
  & & criteria for feedback & \cite{yuan2024llmcrit_W15} \\ 
  & & principles for ideation & \cite{lee2024generating_D7} \\ 
  & & conceptual moves for ideation & \cite{heyman2024supermind_G2} \\ 
  \cmidrule{1-4}

  \multirow{5}{*}{\shortstack{\textbf{Decomposition} \\ \textbf{\& Composition}}} & \textbf{Decomposition} & task-decompose & \cite{wang2023task_D8}, \cite{qian2024shape_M5} \\
  \cmidrule{2-4}  
  & \multirow{4}{*}{\textbf{Composition}} & recombines and mutates & \cite{lanzi2023chatgpt_D6} \\
  & & conceptual blending & \cite{wang2023popblends_M2} \\
  & & combinational creativity & \cite{chen2024foundation_M8} \\
  & & recomposing facts & \cite{pu2024ideasynth_S1}, \cite{radensky2024scideator_S3} \\
  \cmidrule{1-4}  

  \multirow{4}{*}{\shortstack{\textbf{Analogical} \\  \textbf{Thinking \phantom{1}}}} & \multirow{4}{*}{\textbf{Analogical Thinking}}
  & analogy mining and generation & \cite{bhavya2023cam_G16} \\
  & & analogy design heuristic & \cite{veloso2024forming_D12} \\  
  & & recursion analogies generation & \cite{bernstein2024like_G11} \\
  & & bioinspired analogies and tools & \cite{kang2024biospark_G14}, \cite{chen2024asknaturenet_G10}  \\
  \cmidrule{1-4}

  \multirow{7}{*}{\shortstack{\textbf{Multi- \phantom{11111}} \\ \textbf{Dimensional} \\ \textbf{Thinking \phantom{111}}}} & \textbf{Design Space} & develop design space & \cite{suh2024luminate_G13}, \cite{cai2023designaid_M1} \\
  \cmidrule{2-4}  
  & \multirow{2}{*}{\textbf{Multiple Variations}}
  & idea variation tool & \cite{di2022idea_G4} \\
  & & multiple writing variations & \cite{reza2024abscribe_W13} \\
  \cmidrule{2-4} 
  & \multirow{4}{*}{\textbf{Persona-Based Thinking}}
  & user personas for product design & \cite{schuller2024generating_G9} \\
  & & personas for project design & \cite{kocaballi2023conversational_D1} \\  
  & & personas for research ideation & \cite{liu2024personaflow_S2} \\
  & & personas for writing feedback & \cite{benharrak2024writer_W14} \\
  \cmidrule{1-4}

  \textbf{Information-driven Thinking} & \textbf{Information Recommendation}
  & real-time information recommendation & \cite{blazevic2024real_G15} \\
  \cmidrule{1-4}

  \multirow{2}{*}[-0.3\baselineskip]{\shortstack{\textbf{Scenario-Based} \\  \textbf{Thinking \phantom{11111}}}} & \textbf{AR-based Scenarios}
  & body movement for brainstorming & \cite{aikawa2023introducing_G8} \\
  \cmidrule{2-4}
  & \textbf{Driving-Based scenarios} & context-aware for creative incubation  & \cite{paredes2024creative_G17} \\
  \cmidrule{1-4}

  \multirow{5}{*}{\shortstack{\textbf{Collaborative} \\ \textbf{Ideation \phantom{111}}}} & \multirow{5}{*}{\textbf{Group Ideation}}
  & general group ideation & \cite{gonzalez2024collaborative_G7}, \cite{blazevic2024real_G15}, \cite{bilgram2023accelerating_D2}, \cite{lanzi2023chatgpt_D6}, \cite{ege2024chatgpt_D10}, \cite{junior2023chatgpt_D11} \\
  & & conversational deliberation & \cite{rosenberg2023conversational_G6} \\
  & & brainstorming & \cite{aikawa2023introducing_G8} \\
  & & brainwriting & \cite{shaer2024ai_G1} \\  
    
  \bottomrule
\end{longtable}

Structured thinking methods encompass the thinking chain method, which includes generating chains of thought \cite{lee2024prompt_G18,hou2024c2ideas_D13}, graphs of thoughts \cite{besta2024graph_G12}, causal map \cite{huang2023causalmapper_G5}, and mind maps \cite{liu2024personaflow_S2,zhang2023visar_W4}. Additionally, structured templates are utilized to guide ideation, and include those used for reflection \cite{xu2024jamplate_G3}, criteria for feedback \cite{yuan2024llmcrit_W15}, principles for ideation \cite{lee2024generating_D7}, and conceptual moves for ideation \cite{heyman2024supermind_G2}.

Decomposition and composition methods involve breaking down tasks into smaller components \cite{wang2023task_D8} and recombining or mutating ideas through various techniques, such as conceptual blending \cite{wang2023popblends_M2}, combinational creativity \cite{chen2024foundation_M8}, and recomposing facts \cite{pu2024ideasynth_S1,radensky2024scideator_S3}.

Analogical thinking methods focus on generating and leveraging analogies for ideation, including analogy mining and generation \cite{bhavya2023cam_G16}, analogy design heuristics \cite{veloso2024forming_D12}, recursion analogies \cite{bernstein2024like_G11}, and bioinspired analogies and tools \cite{kang2024biospark_G14,chen2024asknaturenet_G10}.

Multi-dimensional thinking methods encompass developing design spaces \cite{suh2024luminate_G13,cai2023designaid_M1}, generating multiple variations of ideas \cite{di2022idea_G4,reza2024abscribe_W13}, and employing persona-based thinking for various purposes, such as product design \cite{schuller2024generating_G9}, project design \cite{kocaballi2023conversational_D1}, research ideation \cite{liu2024personaflow_S2}, and writing feedback \cite{benharrak2024writer_W14}.

Information-driven thinking involves the use of real-time information recommendation to support ideation \cite{blazevic2024real_G15}. Scenario-based thinking methods include using AR-based scenarios for body movement brainstorming \cite{aikawa2023introducing_G8} and driving-based scenarios for context-aware creative incubation \cite{paredes2024creative_G17}.

Lastly, collaborative ideation methods focus on group ideation activities, such as general group ideation \cite{gonzalez2024collaborative_G7,blazevic2024real_G15,bilgram2023accelerating_D2,lanzi2023chatgpt_D6,ege2024chatgpt_D10,junior2023chatgpt_D11}, conversational deliberation \cite{rosenberg2023conversational_G6}, brainstorming \cite{aikawa2023introducing_G8}, and brainwriting \cite{shaer2024ai_G1}.

\newpage
\section{The Hourglass Ideation Framework for LLM-assisted Ideation}

\begin{figure}[ht!]
  \centering
  \includegraphics[width=1\textwidth]{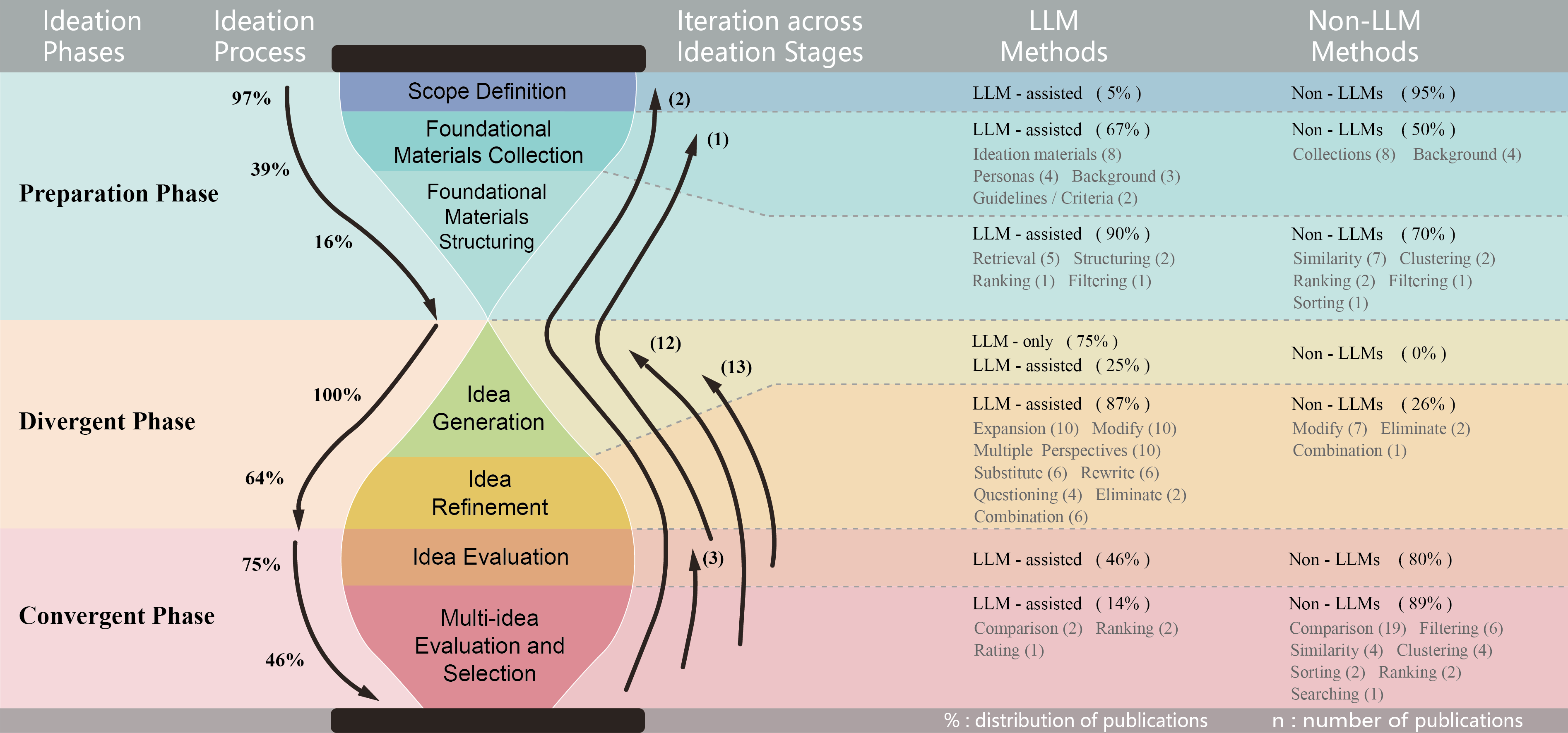}
  \caption{The Hourglass Ideation Framework for LLM-assisted Ideation. The hourglass shape of the framework visualizes the divergent and convergent dynamics of the ideation process, which consists of three overarching phases: preparation (blue), divergent (orange), and convergent (red). Within these phases, there are seven key stages: scope definition, foundational materials collection, foundational materials structuring, idea generation, idea refinement, idea evaluation, and multi-idea evaluation and selection. The framework emphasizes the iterative nature of the ideation process, with arrows indicating potential loops between stages. The bars show the involvement of LLMs and non-LLMs methods in each stage. The percentages and numbers represent the proportion and quantity of reviewed papers, respectively.}
  \label{Framework}
\end{figure}

To understand the roles and contributions of large language models (LLMs) in supporting ideation, we propose the Hourglass Ideation Framework (Figure \ref{Framework}), an iterative model that captures the key stages and interactions involved in leveraging LLMs for creative ideation. The framework draws inspiration from the structuring of ideation factors proposed by Gerlach \cite{gerlach2017idea}, but is derived from our comprehensive survey of LLM-assisted ideation literature. It adapts and extends the existing model to align with the unique characteristics and requirements of LLM-based ideation processes.

The Hourglass Ideation Framework (Figure \ref{Framework}) comprises seven key stages which are organized into three overarching phases: preparation, divergent, and convergent. The preparation phase, represented by the narrow top of the hourglass, establishes the groundwork for ideation by defining objectives and curating relevant resources. The divergent phase, forming the expanding middle section, focuses on generating and exploring a wide range of ideas, while the convergent phase, mirroring the narrow bottom, narrows down and refines the most promising concepts.

A distinguishing feature of the Hourglass Ideation Framework is its emphasis on the iterative nature of LLM-assisted ideation. The framework recognizes that the ideation process is often non-linear, involving multiple interactions and revisits to preceding stages. This iterative approach, represented by the arrows between stages, allows for continuous refinement as the process moves back and forth, enabling ideators to incorporate feedback, new insights, and evolving perspectives.

By providing a structured yet flexible representation of the LLM-assisted ideation process, the Hourglass Ideation Framework serves as a valuable tool for researchers, designers, and practitioners. It offers a common language and conceptual foundation for understanding and discussing the complex interactions between humans and LLMs in creative ideation tasks. The framework can guide the design and evaluation of LLM-assisted ideation systems, inform the development of new methodologies and techniques, and facilitate comparative analyses across different domains and applications.

\newpage
\subsection{An Iterative Framework for LLM Assistance}

\textbf{Preparation Phase}

\begin{itemize}
\item [1.]\textit{Scope Definition} - 
This stage is a crucial starting point for LLM-assisted ideation, laying the foundation for the entire creative process. It involves clearly defining the parameters, objectives, and constraints of the ideation task, in order to guide the subsequent stages effectively. The ideation scope can be determined or enhanced by the organizers, researchers, participants themselves, or LLMs, depending on the study's goals and research questions.

\item [2.]\textit{Foundational Materials Collection} -
This materials collection stage focuses on curating and synthesizing relevant foundational materials to inform and stimulate the subsequent ideation processes. This content can be categorized into five main types: background information, ideation materials, data collections, guidelines and criteria, and personas. LLMs can contribute to this stage by generating guidelines, criteria, personas, and summaries of ideation materials and scholarly literature.

\item [3.]\textit{Foundational Materials Structuring} -
This materials structuring stage focuses on organizing and synthesizing the diverse texts and data sources gathered during the previous stage. It plays a pivotal role in the LLM-assisted ideation process, as it helps researchers systematically analyse and refine the input resources, facilitating their effective utilization in subsequent ideation stages. LLMs contribute to this process primarily through text retrieval, structured encoding, ranking, and filtering.
\end{itemize}

\textbf{Divergent Phase}

\begin{itemize}
\item [3.]\textit{Idea Generation} - 
This stage lies at the heart of LLM-assisted ideation, where novel ideas and concepts are produced through the interaction of human creativity and LLM capabilities. The emphasis at this stage is on generating a large quantity and diverse range of ideas, prioritizing creativity and novelty over immediate evaluation or feasibility. This stage may rely entirely on LLMs to generate ideas or employ a combination of LLM and human input.  

\item [4.]\textit{Idea Refinement} -
The idea refinement stage focuses on enhancing the initial ideas to improve their originality, relevance, and impact. LLMs contribute to this process through methods such as expansion, modification, substitution, rewriting, elimination, multiple perspectives, questioning, and combination.
\end{itemize}

\textbf{Convergent Phase}

\begin{itemize}
\item [5.]\textit{Idea Evaluation} - 
The stage considers each idea in turn or in isolation. It plays a critical role in assessing the quality, feasibility, and potential impact of generated ideas based on predefined criteria and feedback from various perspectives. This stage helps identify the most promising ideas for further development and implementation, ensuring that the LLM-assisted ideation process yields practical and valuable outcomes. LLMs contribute to this stage through methods such as providing feedback, scoring ideas, assessing novelty, and evaluating ideas using general or specific criteria.

\item [6.]\textit{Multiple Idea Evaluation and Selection} - 
The multiple idea evaluation and selection stage focuses on assessing and selecting the most promising ideas from the pool of evaluated concepts. This stage is crucial for identifying the most creative and valuable ideas generated during the ideation process. LLMs contribute to this stage through methods such as comparing, ranking, and rating multiple ideas to facilitate the convergent process. 

\end{itemize}

Thus this framework allows us both to structure our review and describe the complex interactions between humans and AI during the ideation process. The insights from the ideation framework could also guide the design of ideation processes for specific problems.

\subsection{Overview of LLM use in Ideation Stages}

From the 61 papers we reviewed, Table \ref{tb:Review} presents an overview of how LLMs have been used within different ideation stages. Note that the techniques described cover both LLM($\blacksquare$) and non-LLM($\square$) methods. Stages annotated with both symbols ($\square$$\blacksquare$) indicate that a stage is performed jointly using LLM and non-LLM techniques. 

Note that spreadsheet versions of the tables presented in this review are provided separately \footnote[1]{Resources are available at  \url{https://doi.org/10.6084/m9.figshare.28440182}}.

\begin{table}[H]
  \centering
  \normalsize
  \caption{Large language models' (LLMs) contributions to the critical ideation stages. Includes LLM($\blacksquare$), Non-LLM($\square$) and both($\square$$\blacksquare$) methods.}
  \label{tb:Review}
  \resizebox{\textwidth}{!}{
  \begin{tabular}{>{\raggedright\arraybackslash}m{0.8cm} >{\centering\arraybackslash}m{2cm} >
  {\centering\arraybackslash}m{1cm} >
  {\centering\arraybackslash}m{3cm} >{\centering\arraybackslash}m{3cm} >
  {\centering\arraybackslash}m{3cm} >  
  {\centering\arraybackslash}m{3cm} >{\centering\arraybackslash}m{3cm} >
  {\centering\arraybackslash}m{3cm} >
  {\centering\arraybackslash}m{3cm} >  {\raggedright\arraybackslash}m{2cm}}
  \toprule
  & & & & \textbf{2.Foundational} & \textbf{3.Foundational} & & & & \textbf{7.Multiple Idea} &  \\
  \textbf{Ref.} & \textbf{Domain} & \textbf{Index} & \textbf{1.Scope Definition} & \textbf{Materials} & \textbf{Materials} & \textbf{4.Idea Generation} & \textbf{5.Idea Refinement} & \textbf{6.Idea Evaluation} & \textbf{Evaluation} & \textbf{Iteration} \\
  & & & & \textbf{Collection} & \textbf{Structuring} & & & & \textbf{and Selection} &  \\
  \midrule
  \cite{shaer2024ai_G1} & General & G1 & $\square\phantom{\blacksquare}$ & - & - & $\square\blacksquare$ & $\square\phantom{\blacksquare}$ & $\square\blacksquare$ & $\square\phantom{\blacksquare}$ & - \\
  \cite{heyman2024supermind_G2} & General & G2 & $\square\phantom{\blacksquare}$ & - & - & $\phantom{\square}\blacksquare$ & - & $\square\phantom{\blacksquare}$ & - & 6 -→ 4 \\
  \cite{xu2024jamplate_G3} & General & G3 & $\square\phantom{\blacksquare}$ & $\square\blacksquare$ & - & $\phantom{\square}\blacksquare$ & $\square\blacksquare$ & $\square\blacksquare$ & $\phantom{\square}\blacksquare$ & 7 -→ 4 / 5 \\  
  \cite{di2022idea_G4} & General & G4 & $\square\phantom{\blacksquare}$ & - & - & $\square\blacksquare$ & $\phantom{\square}\blacksquare$ & $\square\phantom{\blacksquare}$ & $\square\phantom{\blacksquare}$ & - \\ 
  \cite{huang2023causalmapper_G5} & General & G5 & $\square\phantom{\blacksquare}$ & - & - & $\phantom{\square}\blacksquare$ & $\square\phantom{\blacksquare}$ & $\square\phantom{\blacksquare}$ & $\square\phantom{\blacksquare}$ & 7 -→ 6 -→ 4 / 5 \\ 
  \cite{rosenberg2023conversational_G6} & General & G6 & $\square\phantom{\blacksquare}$ & $\phantom{\square}\blacksquare$ & - & $\phantom{\square}\blacksquare$ & - & $\phantom{\square}\blacksquare$ & - & -  \\ 
  \cite{gonzalez2024collaborative_G7} & General & G7 & $\square\phantom{\blacksquare}$ & - & - & $\phantom{\square}\blacksquare$ & $\phantom{\square}\blacksquare$ & $\square\phantom{\blacksquare}$ & $\square\phantom{\blacksquare}$ & 7 -→ 6 -→ 4 / 5 \\   
  \cite{aikawa2023introducing_G8} & General & G8 & - & - & - & $\phantom{\square}\blacksquare$ & $\phantom{\square}\blacksquare$ & $\phantom{\square}\blacksquare$ & - & - \\ 
  \cite{schuller2024generating_G9} & General & G9 & $\square\phantom{\blacksquare}$ & $\phantom{\square}\blacksquare$ & - & $\phantom{\square}\blacksquare$ & - & - & - & - \\ 
  \cite{chen2024asknaturenet_G10} & General & G10 & $\square\phantom{\blacksquare}$ & $\square\phantom{\blacksquare}$ & $\square\blacksquare$ & $\phantom{\square}\blacksquare$ & - & - & - & - \\
  \cite{bernstein2024like_G11} & General & G11 & $\square\phantom{\blacksquare}$ & - & - & $\phantom{\square}\blacksquare$ & - & - & - & - \\
  \cite{besta2024graph_G12} & General & G12 & $\square\phantom{\blacksquare}$ & - & - & $\phantom{\square}\blacksquare$ & $\phantom{\square}\blacksquare$ & $\square\blacksquare$ & $\square\phantom{\blacksquare}$ & 7 -→ 4 / 5 \\
  \cite{suh2024luminate_G13} & General & G13 & $\square\phantom{\blacksquare}$ & - & - & $\square\blacksquare$ & $\square\blacksquare$ & $\square\phantom{\blacksquare}$ & $\square\phantom{\blacksquare}$ & 7 -→ 4 / 5 \\
  \cite{kang2024biospark_G14} & General & G14 & $\square\phantom{\blacksquare}$ & $\square\phantom{\blacksquare}$ & $\square\blacksquare$ & $\phantom{\square}\blacksquare$ & $\phantom{\square}\blacksquare$ & $\square\blacksquare$ & $\phantom{\square}\blacksquare$ & - \\
  \cite{blazevic2024real_G15} & General & G15 & $\square\phantom{\blacksquare}$ & $\square\phantom{\blacksquare}$ & $\square\blacksquare$ & $\phantom{\square}\blacksquare$ & $\phantom{\square}\blacksquare$ & - & $\phantom{\square}\blacksquare$ & - \\
  \cite{bhavya2023cam_G16} & General & G16 & $\square\phantom{\blacksquare}$ & - & - & $\phantom{\square}\blacksquare$ & $\phantom{\square}\blacksquare$ & $\square\blacksquare$ & $\square\blacksquare$ & 7 -→ 4 / 5 \\
  \cite{paredes2024creative_G17} & General & G17 & - & $\square\phantom{\blacksquare}$ & - & $\square\blacksquare$ & - & - & - & -   \\
  \cite{lee2024prompt_G18} & General & G18 & $\square\phantom{\blacksquare}$ & - & - & $\phantom{\square}\blacksquare$ & - & - & - & - \\
  
  \cite{kocaballi2023conversational_D1} & Design & D1 & $\square\phantom{\blacksquare}$ & $\phantom{\square}\blacksquare$ & - & $\phantom{\square}\blacksquare$ & $\phantom{\square}\blacksquare$ & $\phantom{\square}\blacksquare$ & - & 6 -→ 2 \\
  \cite{bilgram2023accelerating_D2} & Design & D2 & $\square\phantom{\blacksquare}$ & $\phantom{\square}\blacksquare$ & - & $\phantom{\square}\blacksquare$ & $\phantom{\square}\blacksquare$ & $\square\phantom{\blacksquare}$ & - & 6 -→ 4 / 5 \\
  \cite{sudhakaran2023prompt_D3} & Design & D3 & $\square\phantom{\blacksquare}$ & - & - & $\phantom{\square}\blacksquare$ & - & $\square\phantom{\blacksquare}$ & - & - \\  
  \cite{todd2023level_D4} & Design & D4 & $\square\phantom{\blacksquare}$ & - & - & $\phantom{\square}\blacksquare$ & - & $\square\phantom{\blacksquare}$ & - & - \\ 
  \cite{dharuman2023protein_D5} & Design & D5 & $\square\phantom{\blacksquare}$ & - & - & $\phantom{\square}\blacksquare$ & - & $\phantom{\square}\blacksquare$ & $\square\phantom{\blacksquare}$ & - \\ 
  \cite{lanzi2023chatgpt_D6} & Design & D6 & $\square\phantom{\blacksquare}$ & - & - & $\square\blacksquare$ & $\phantom{\square}\blacksquare$ & $\square\phantom{\blacksquare}$ & $\square\phantom{\blacksquare}$ & 7 -→ 4 / 5 \\ 
  \cite{lee2024generating_D7} & Design & D7 & $\square\phantom{\blacksquare}$ & $\square\blacksquare$ & $\square\blacksquare$ & $\phantom{\square}\blacksquare$ & - & $\square\phantom{\blacksquare}$ & - & - \\   
  \cite{wang2023task_D8} & Design & D8 & $\square\phantom{\blacksquare}$ & - & - & $\phantom{\square}\blacksquare$ & $\phantom{\square}\blacksquare$ & $\square\phantom{\blacksquare}$ & - & - \\ 
  \cite{lu2024large_D9} & Design & D9 & $\square\phantom{\blacksquare}$ & - & - & $\phantom{\square}\blacksquare$ & - & - & - & - \\ 
  \cite{ege2024chatgpt_D10} & Design & D10 & $\square\phantom{\blacksquare}$ & $\square\phantom{\blacksquare}$ & - & $\phantom{\square}\blacksquare$ & $\phantom{\square}\blacksquare$ & $\phantom{\square}\blacksquare$ & - & 6 -→ 4 / 5 \\
  \cite{junior2023chatgpt_D11} & Design & D11 & $\square\phantom{\blacksquare}$ & $\square\phantom{\blacksquare}$ & - & $\phantom{\square}\blacksquare$ & $\phantom{\square}\blacksquare$ & $\square\blacksquare$ & - & 6 -→ 4 / 5 \\
  \cite{veloso2024forming_D12} & Design & D12 & $\square\phantom{\blacksquare}$ & $\square\blacksquare$ & - & $\phantom{\square}\blacksquare$ & - & - & - & - \\
  \cite{hou2024c2ideas_D13} & Design & D13 & $\square\phantom{\blacksquare}$ & - & - & $\phantom{\square}\blacksquare$ & $\square\phantom{\blacksquare}$ & $\square\phantom{\blacksquare}$ & $\square\phantom{\blacksquare}$ & 6 -→ 4 \\
  \cite{huang2024plantography_D14} & Design & D14 & $\square\phantom{\blacksquare}$ & - & - & $\phantom{\square}\blacksquare$ & $\square\phantom{\blacksquare}$ & $\square\phantom{\blacksquare}$ & - & - \\

  \cite{goodman2022lampost_W1} & Writing & W1 & $\square\phantom{\blacksquare}$ & $\phantom{\square}\blacksquare$ & - & $\square\blacksquare$ & $\phantom{\square}\blacksquare$ & - & $\square\phantom{\blacksquare}$ & - \\
  \cite{kim2024towards_W2} & Writing & W2 & $\square\phantom{\blacksquare}$ & $\phantom{\square}\blacksquare$ & - & $\square\blacksquare$ & $\phantom{\square}\blacksquare$ & $\phantom{\square}\blacksquare$ & - & - \\
  \cite{chung2022talebrush_W3} & Writing & W3 & $\square\phantom{\blacksquare}$ & - & - & $\square\blacksquare$ & $\phantom{\square}\blacksquare$ & $\square\phantom{\blacksquare}$ &  $\square\phantom{\blacksquare}$ & 7 -→ 4 \\  
  \cite{zhang2023visar_W4} & Writing & W4 & $\square\blacksquare$ & - & - & $\phantom{\square}\blacksquare$ & $\phantom{\square}\blacksquare$ & $\square\blacksquare$ & $\square\phantom{\blacksquare}$ & 7 -→ 4 / 5 \\ 
  \cite{yuan2022wordcraft_W5} & Writing & W5 & $\square\phantom{\blacksquare}$ & - & - & $\square\blacksquare$ & $\phantom{\square}\blacksquare$ & $\square\phantom{\blacksquare}$ & $\square\phantom{\blacksquare}$ & - \\ 
  \cite{petridis2023anglekindling_W6} & Writing & W6 & $\square\phantom{\blacksquare}$ & - & - & $\phantom{\square}\blacksquare$ & $\phantom{\square}\blacksquare$ & - & -  & - \\ 
  \cite{ghajargar2022redhead_W7} & Writing & W7 & $\square\phantom{\blacksquare}$ & - & - & $\phantom{\square}\blacksquare$ & $\phantom{\square}\blacksquare$ & - & - & - \\   
  \cite{wan2024felt_W8} & Writing & W8 & $\square\phantom{\blacksquare}$ & - & - & $\square\blacksquare$ & - & - & - & - \\ 
  \cite{kim2023cells_W9} & Writing & W9 & $\square\phantom{\blacksquare}$ & - & - & $\phantom{\square}\blacksquare$ & - & $\square\phantom{\blacksquare}$ & $\square\phantom{\blacksquare}$ & 7 -→ 4 \\ 
  \cite{goldi2024intelligent_W10} & Writing & W10 & $\square\phantom{\blacksquare}$ & - & - & $\square\blacksquare$ & - & $\square\blacksquare$ & $\square\phantom{\blacksquare}$ & - \\
  \cite{li2024mystery_W11} & Writing & W11 & $\square\phantom{\blacksquare}$ & - & - & $\phantom{\square}\blacksquare$ & - & $\square\phantom{\blacksquare}$ & - & 6 -→ 4 \\
  \cite{pividori2024publishing_W12} & Writing & W12 & $\square\phantom{\blacksquare}$ & - & - & $\phantom{\square}\blacksquare$ & - & - & - & - \\
  \cite{reza2024abscribe_W13} & Writing & W13 & $\square\phantom{\blacksquare}$ & - & - & $\square\blacksquare$ & $\phantom{\square}\blacksquare$ & $\square\phantom{\blacksquare}$ & $\square\phantom{\blacksquare}$ & 7 -→ 4 / 5 \\
  \cite{benharrak2024writer_W14} & Writing & W14 & $\square\phantom{\blacksquare}$ & $\phantom{\square}\blacksquare$ & - & $\square\blacksquare$ & $\phantom{\square}\blacksquare$ & $\square\blacksquare$ & $\square\phantom{\blacksquare}$ & 7 -→ 4 / 5 \\
  \cite{yuan2024llmcrit_W15} & Writing & W15 & $\square\phantom{\blacksquare}$ & $\square\blacksquare$ & - & $\phantom{\square}\blacksquare$ & $\square\phantom{\blacksquare}$ & $\square\blacksquare$ & - & - \\

  \cite{pu2024ideasynth_S1} & Science & S1 & $\square\blacksquare$ & $\square\blacksquare$ & $\phantom{\square}\blacksquare$ & $\phantom{\square}\blacksquare$ & $\square\blacksquare$ & $\square\blacksquare$ & $\square\phantom{\blacksquare}$ & 7 -→ 1 \\
  \cite{liu2024personaflow_S2} & Science & S2 & $\square\blacksquare$ & $\square\blacksquare$ & $\square\phantom{\blacksquare}$ & $\phantom{\square}\blacksquare$ & $\square\blacksquare$ & $\phantom{\square}\blacksquare$ & $\square\phantom{\blacksquare}$ & 7 -→ 1 \\
  \cite{radensky2024scideator_S3} & Science & S3 & $\square\phantom{\blacksquare}$ & $\square\blacksquare$ & $\phantom{\square}\blacksquare$ & $\square\blacksquare$ & $\square\blacksquare$ & $\phantom{\square}\blacksquare$ & $\square\phantom{\blacksquare}$ & 7 -→ 4 / 5 \\  
  \cite{banker2024machine_S4} & Science & S4 & $\square\phantom{\blacksquare}$ & - & - & $\square\blacksquare$ & - & - & - & -  \\ 
  \cite{chen2024use_S5} & Science & S5 & $\square\phantom{\blacksquare}$ & $\square\phantom{\blacksquare}$ & - & $\phantom{\square}\blacksquare$ & - & $\square\phantom{\blacksquare}$ & - & 6 -→ 4 \\ 
  \cite{lee2024gpt_S6} & Science & S6 & $\square\phantom{\blacksquare}$ & - & - & $\phantom{\square}\blacksquare$ & $\phantom{\square}\blacksquare$ & $\phantom{\square}\blacksquare$ & - & 6 -→ 4 / 5 \\  
  \cite{cai2023designaid_M1} & Multimedia & M1 & $\square\phantom{\blacksquare}$ & - & - & $\phantom{\square}\blacksquare$ & $\phantom{\square}\blacksquare$ & $\square\phantom{\blacksquare}$ & $\square\phantom{\blacksquare}$ & 7 -→ 4 / 5 \\
  \cite{wang2023popblends_M2} & Multimedia & M2 & $\square\phantom{\blacksquare}$ & - & - & $\phantom{\square}\blacksquare$ & $\phantom{\square}\blacksquare$ & $\square\phantom{\blacksquare}$ & $\square\phantom{\blacksquare}$ & - \\
  \cite{brade2023promptify_M3} & Multimedia & M3 & $\square\phantom{\blacksquare}$ & - & - & $\phantom{\square}\blacksquare$ & $\phantom{\square}\blacksquare$ & $\square\phantom{\blacksquare}$ & $\square\phantom{\blacksquare}$ & 7 -→ 6 -→ 4 / 5 \\  
  \cite{wang2024lave_M4} & Multimedia & M4 & $\square\phantom{\blacksquare}$ & $\phantom{\square}\blacksquare$ & $\square\blacksquare$ & $\phantom{\square}\blacksquare$ & $\phantom{\square}\blacksquare$ & $\square\phantom{\blacksquare}$ & - & - \\ 
  \cite{qian2024shape_M5} & Multimedia & M5 & $\square\phantom{\blacksquare}$ & $\square\phantom{\blacksquare}$ & $\square\blacksquare$ & $\phantom{\square}\blacksquare$ & $\phantom{\square}\blacksquare$ & $\square\phantom{\blacksquare}$ & - & 6 -→ 4 / 5 \\ 
  \cite{wang2023script_M6} & Multimedia & M6 & $\square\phantom{\blacksquare}$ & $\phantom{\square}\blacksquare$ & $\phantom{\square}\blacksquare$  & $\phantom{\square}\blacksquare$ & - & $\square\phantom{\blacksquare}$ & - & - \\ 
  \cite{wu2023iconshop_M7} & Multimedia & M7 & $\square\phantom{\blacksquare}$ & - & - & $\phantom{\square}\blacksquare$ & - & - & - & - \\   
  \cite{chen2024foundation_M8} & Multimedia & M8 & $\square\phantom{\blacksquare}$ & - & - & $\phantom{\square}\blacksquare$ & $\phantom{\square}\blacksquare$ & $\square\blacksquare$ & $\square\phantom{\blacksquare}$ & 6 -→ 4 / 5 \\

  \bottomrule
  \end{tabular}
  }
\end{table}

Table \ref{tb:Review} is primarily intended to allow readers to focus on articles that discuss particular ideation stages (or combinations of such stages) and to unify many of the results presented in this paper.

The table reveals that the idea generation stage is the most frequently explored stage in LLM-assisted ideation activities, followed by the idea refinement stage. This suggests that researchers and practitioners are primarily focused on leveraging the generative and refinement capabilities of LLMs to enhance the creative divergent process.

The preparation phase, including scope definition, foundational materials collection, and foundational materials structuring, are less frequently investigated compared to the divergent phase. However, among these preparation stages, the foundational materials collection and structuring stages see a higher proportion of LLM involvement, indicating the potential for LLMs to support the curation, organization, and processing of relevant data.

In the convergent phase, the idea evaluation stage is more commonly explored than the multi-idea evaluation and selection stage. This suggests that researchers are interested in utilizing LLMs to assess the quality and potential of individual ideas, rather than using LLMs for comparing and selecting the most suitable ideas.

Overall, the table highlights the varying degrees of LLM use across the different stages, with the divergent phase seeing the highest levels of LLM involvement, followed by the preparation phase, and finally, the convergent phase. This pattern indicates that LLMs are primarily employed to generate and refine ideas, while their use in the evaluation and selection processes is still an emerging area of research.

\subsection{The use of LLMs within the different Ideation Stages}

\subsubsection{Scope Definition }

This stage is a critical starting point for LLM-assisted ideation, laying the foundation for the entire creative process. It involves clearly defining the parameters, objectives, and constraints of the ideation task to guide the subsequent stages effectively. Our analysis reveals that 59 out of the 61 studies (97\%) included a scope definition stage, underscoring its significance in the ideation workflow.

In 53 out of 59 studies, the ideation scope was determined by the organizers or researchers, ensuring alignment with the study's goals and research questions. However, in 3 studies related to writing tasks \cite{goodman2022lampost_W1,kim2024towards_W2,benharrak2024writer_W14}, participants were allowed to define their own scope by providing specific writing prompts or objectives. This approach highlights the potential for user-driven scope definition in certain ideation contexts.

Notably, only 3 studies (5\%) employed LLMs to assist in the scope definition process. In study \cite{pu2024ideasynth_S1}, the LLM helped refine the initial research problem by generating clarifying questions or proposing alternative scopes based on relevant literature from the pre-ideation foundations. Similarly, study \cite{liu2024personaflow_S2} utilized an LLM to modify the user's initial research question into a revised problem statement. In study \cite{zhang2023visar_W4}, the LLM provided hierarchical writing goal suggestions to guide the scope definition for argumentative writing tasks.

These examples demonstrate the potential for LLMs to play a more active role in assisting users to define and refine the scope of their ideation tasks. By generating probing questions, suggesting alternative perspectives, and providing structured guidance, LLMs can help users articulate their creative objectives more clearly and comprehensively. Also, this is surprising given the importance of this stage and its value for use in the evaluation stage. However, the limited number of studies employing LLMs in this stage suggests that this area is still underexplored and presents opportunities for future research.

\subsubsection{Foundational Materials Collection}

This stage focuses on curating and synthesizing relevant background information, inspiration materials, datasets, guidelines, and personas to inform and stimulate the subsequent ideation processes. Our review found that 24 out of the 61 studies (39\%) included a foundational materials collection stage, indicating its important, but not universally essential, role in LLM-assisted ideation workflows.

The foundational materials content can be categorized into five main types: 

1) Background information, such as existing solutions \cite{junior2023chatgpt_D11,veloso2024forming_D12}, market conditions \cite{bilgram2023accelerating_D2}, competitor information \cite{xu2024jamplate_G3}, and scenario-based context \cite{paredes2024creative_G17}; 

2) Ideation materials, including textual prompts for writing inspiration \cite{goodman2022lampost_W1,kim2024towards_W2}, video clips \cite{wang2024lave_M4}, audio recordings of group discussions \cite{rosenberg2023conversational_G6}, story elements for picture book creation \cite{wang2023script_M6}, and scholarly literature for exploring research ideas \cite{pu2024ideasynth_S1,liu2024personaflow_S2,radensky2024scideator_S3};

3) Data collections, such as biology databases \cite{chen2024asknaturenet_G10,kang2024biospark_G14}, patent repositories \cite{lee2024generating_D7}, specialized document corpora \cite{yuan2024llmcrit_W15}, engineering prototype datasets \cite{ege2024chatgpt_D10}, custom code libraries \cite{qian2024shape_M5}, and comprehensive research resources \cite{blazevic2024real_G15,chen2024use_S5};

4) Guidelines and criteria, including ecological design principles \cite{lee2024generating_D7}, authoritative references \cite{yuan2024llmcrit_W15}, and standards for generating creative feedback \cite{yuan2024llmcrit_W15};

5) Personas, representing diverse perspectives such as product testers \cite{schuller2024generating_G9}, designers, users, and products \cite{kocaballi2023conversational_D1}, as well as customized roles for providing writing feedback \cite{benharrak2024writer_W14} and evaluating scientific ideas \cite{liu2024personaflow_S2}.

Among the studies that incorporated the foundational materials collection stage, 16 (67\%) utilized LLMs to help generate or summarize foundational content. They contribute to this stage in two ways. First, they can generate background \cite{xu2024jamplate_G3,bilgram2023accelerating_D2,veloso2024forming_D12}, guidelines \cite{lee2024generating_D7}, criteria \cite{yuan2024llmcrit_W15}, and personas \cite{schuller2024generating_G9,kocaballi2023conversational_D1,benharrak2024writer_W14,liu2024personaflow_S2} that guide and inspire the ideation process. Second, LLMs can summarize and synthesize ideation materials \cite{rosenberg2023conversational_G6,goodman2022lampost_W1,kim2024towards_W2,wang2024lave_M4,wang2023script_M6} and scholarly literature \cite{pu2024ideasynth_S1,liu2024personaflow_S2,radensky2024scideator_S3}, extracting key insights to inform the creative process.

These findings suggest that LLMs can play a valuable role in curating, generating, and processing pre-ideation content, helping to establish a rich and diverse foundation for creative exploration. By leveraging their ability to analyse and summarize large volumes of data, LLMs can surface relevant information, identify patterns and connections, and distill complex materials into actionable insights.

However, the relatively low proportion of studies incorporating foundational materials collection also indicates potential challenges and limitations. Curating high-quality, domain-specific datasets and resources may require significant effort and expertise, particularly in specialized or emerging fields. Additionally, the effectiveness of LLM-generated guidelines, personas, and summaries will depend on the quality and relevance of the training data, as well as the ability to fine-tune the models for specific ideation contexts.

\subsubsection{Foundational Materials Structuring}

This stage plays a crucial role in organizing and processing the diverse data sources gathered during the preceding stage. Our review found that 10 out of the 61 studies (16\%) included a foundational materials structuring stage, accounting for 42\% of the 24 studies that incorporated a foundational materials collection stage. This highlights its significance in preparing the collected data for effective utilization in the ideation process. Notably, 9 out of the 10 studies (90\%) that performed structuring of the foundation materials employed LLM-based methods, highlighting the significant advantages of LLMs in processing large volumes of textual data.

The primary objective of this stage is to converge and refine the various textual and data resources, with ideation materials and text collections accounting for 81\% of the processed data. This finding emphasizes the importance of carefully curating and filtering large text corpora to enhance the effectiveness of LLMs in leveraging these resources during the subsequent ideation stages.

\begin{longtable}{>{\raggedright\arraybackslash}p{2.5cm} >
  {\raggedright\arraybackslash}p{3cm} >
  {\raggedright\arraybackslash}p{3cm} >
  {\raggedright\arraybackslash}p{4cm} }
  \caption{Foundational Materials Structuring Methods} \label{tb:Foundational Materials Structuring} \\ 
  \toprule
  \textbf{Ideation Stage} & \textbf{Technique} & \textbf{Method} & \textbf{Ref.} \\
  \midrule
  \endfirsthead
  \\
  \\
  \toprule
  \textbf{Ideation Stage} & \textbf{Technique} & \textbf{Method} & \textbf{Ref.} \\
  \midrule
  \endhead
  \multirow{10}{*}{\shortstack{\textbf{Foundational} \\ \textbf{Materials \phantom{xxx}} \\ \textbf{Structuring \phantom{x}}}} & \multirow{5}{*}{\textbf{Non-LLMs}} & Similarity & \cite{chen2024asknaturenet_G10}, \cite{kang2024biospark_G14}, \cite{blazevic2024real_G15}, \cite{lee2024generating_D7}, \cite{wang2024lave_M4}, \cite{qian2024shape_M5}, \cite{liu2024personaflow_S2} \\
  & & Clustering & \cite{lee2024generating_D7}, \cite{kang2024biospark_G14} \\ 
  & & Ranking & \cite{blazevic2024real_G15}, \cite{lee2024generating_D7} \\  
  & & Filtering & \cite{lee2024generating_D7} \\  
  & & Sorting & \cite{kang2024biospark_G14} \\
  \cmidrule{2-4}  
  & \multirow{4}{*}{\textbf{LLMs}} & Retrieval & \cite{pu2024ideasynth_S1}, \cite{radensky2024scideator_S3}, \cite{blazevic2024real_G15}, \cite{qian2024shape_M5}, \cite{lee2024generating_D7}  \\
  & & Structuring & \cite{chen2024asknaturenet_G10}, \cite{kang2024biospark_G14} \\
  & & Ranking & \cite{wang2024lave_M4} \\
  & & Filtering & \cite{wang2023script_M6} \\
  
  \bottomrule
\end{longtable}

As shown in Table \ref{tb:Foundational Materials Structuring}, non-LLM techniques employed in the foundational materials structuring stage include:

1) Similarity calculation: This method involves measuring the semantic relatedness between textual elements, such as documents or concepts, to identify patterns and connections within the collected materials. Similarity metrics, such as cosine similarity \cite{chen2024asknaturenet_G10,lee2024generating_D7,wang2024lave_M4,liu2024personaflow_S2} and Euclidean distance \cite{lee2024generating_D7}, were used to compute the similarity between embeddings of textual data.

2) Clustering: Clustering methods were used to group similar textual elements together based on their semantic or structural properties. Study \cite{lee2024generating_D7} applied text clustering to group semantically similar guidelines generated by LLMs, making them more accessible and user-friendly.

3) Ranking: Non-LLM ranking techniques were applied to prioritize and order the collected materials based on their relevance or importance to the ideation context. For example, the study \cite{blazevic2024real_G15} employed a combination of topic modeling and similarity algorithms to rank scientific publications.

4) Filtering: Filtering methods were used to refine the collected materials by removing irrelevant or redundant information. Study \cite{lee2024generating_D7} applied regular expressions and predefined lists to filter out inaccurate principles generated by LLMs.

5) Sorting: Sorting techniques were employed to arrange the collected materials in a specific order based on predefined criteria. For instance, study \cite{kang2024biospark_G14} sorted taxonomic ranks in an increasing order based on the number of immediate children nodes to identify sparsely populated branches for expansion.

LLMs contribute to the foundational materials structuring stage through four primary approaches:

1) Text retrieval based on semantic similarity: Several studies employed LLMs to identify and retrieve relevant text segments from large corpora by calculating semantic similarity between the query and the stored embeddings \cite{pu2024ideasynth_S1,radensky2024scideator_S3,blazevic2024real_G15,qian2024shape_M5,lee2024generating_D7}. This approach, known as Retrieval-Augmented Generation (RAG), is a cutting-edge technique that combines the strengths of LLMs and information retrieval systems to generate more accurate and relevant outputs.

2) Structured encoding of text data: LLMs were utilized to encode textual data into structured formats, such as knowledge graphs or taxonomic hierarchies \cite{chen2024asknaturenet_G10,kang2024biospark_G14}. This structured representation facilitates the exploration and analysis of complex relationships within the data, enhancing the ideation process.

3) Ranking: LLMs were employed to rank the relevance of video narrations based on their semantic similarity to a given query \cite{wang2024lave_M4}. This ranking mechanism enables users to quickly identify the most pertinent content for their ideation needs.

4) Filtering data information: LLMs were used to filter and refine textual data, such as extracting relevant keywords and descriptions from story scripts \cite{wang2023script_M6}. This filtering process helps to distill the most valuable information from the collected materials, ensuring a more focused and effective ideation process.

However, our review identified a lack of research exploring the use of LLMs for similarity calculation and clustering in the stage of foundational materials structuring. These techniques have been widely used in non-LLM methods, as demonstrated by several studies \cite{chen2024asknaturenet_G10,kang2024biospark_G14,lee2024generating_D7,liu2024personaflow_S2}. The absence of LLM-based similarity and clustering approaches presents an opportunity for future research to investigate how LLMs can be leveraged to enhance the effectiveness and efficiency of these structuring techniques. By incorporating LLMs' semantic understanding capabilities, researchers may be able to develop more sophisticated and context-aware methods for grouping and organizing foundational materials.

Another notable observation is that none of the reviewed studies considered visualizing structured foundational materials, such as using bubble or tree maps to represent topic models and organized information. Visualizing the structured data could provide ideators with a more intuitive and accessible way to explore and interact with the organized materials, potentially fostering new insights and connections. Future research could investigate the integration of data visualization techniques with LLM-based foundational materials structuring to further enhance the ideation process.

\subsubsection{Idea Generation}

The idea generation stage lies at the heart of the LLM-assisted ideation process, where novel ideas and concepts are produced through the interaction of human creativity and LLM capabilities. Our analysis reveals that all 61 studies (100\%) included an idea generation stage, confirming its essential role in LLM-assisted ideation. The prevalence of this stage highlights the effectiveness of LLMs in producing novel and relevant ideas, either independently or in collaboration with human creators.

Among the 61 studies, 46 (75\%) relied entirely on LLMs to generate ideas throughout the ideation activity, without any human input during the idea generation process. Of these LLM-only studies, 42 used the LLM-generated ideas directly in the ideation activity, while 4 studies allowed users to modify and iterate on the LLM-generated ideas. The high proportion of studies relying solely on LLMs for idea generation demonstrates the capability of these models to generate creative and applicable ideas independently. 

In contrast, the remaining 15 studies (25\%) employed a combination of LLM and human input for idea generation, where both the LLM and human creators contributed ideas throughout the ideation process. In 8 of these studies, the LLMs generated ideas that could be used directly in the ideation activity, while 3 studies allowed users to modify and iterate on the LLM-generated ideas. Interestingly, 4 studies used LLMs to generate ideas primarily to stimulate users' creativity rather than to provide directly usable ideas. These tasks were mainly concentrated in the writing domain, suggesting that, for creative writing tasks, people prefer LLMs to provide inspirational prompts rather than dominate the creative process, preserving human agency in the creation process. This finding highlights the potential for LLMs to serve as catalysts for human creativity, rather than simply generating complete solutions.

Despite the promising results, there remain opportunities for further research in optimizing the balance between LLM-generated ideas and human input. Investigating methods for LLMs to generate ideas that stimulate human creativity without dominating the process could help mitigate issues like free riding, where users rely too heavily on the LLM's output. Additionally,  it is important to note that the current literature primarily focuses on adjusting parameters like temperature to regulate the innovativeness of LLM-generated ideas, with no studies exploring methods to flexibly adjust the degree to which LLMs provide directly usable ideas versus ideas that stimulate human creativity during ideation activities.

\subsubsection{Idea Refinement}

The idea refinement stage plays a crucial role in the divergent phase of LLM-assisted ideation processes, where initial ideas are expanded and refined through various methods to make them more comprehensive and sophisticated. Among the 61 studies on LLM-assisted ideation activities reviewed in this survey, 39 studies (64\%) used an idea refinement stage, highlighting its significance for exploring idea creativity and optimization in LLM-assisted ideation activities.

Of the 39 studies that included the idea refinement stage, 34 (87\%) employed LLMs to assist in this process, demonstrating that LLMs have become effective tools for aiding the development of creative ideas.

\begin{longtable}{>{\raggedright\arraybackslash}p{1.8cm} >
  {\raggedright\arraybackslash}p{1.4cm} >{\raggedright\arraybackslash}p{2cm} >
  {\raggedright\arraybackslash}p{2.8cm} >
  {\raggedright\arraybackslash}p{4cm} }
  \caption{Idea Refinement Methods} \label{tb:Idea Refinement} \\ 
  \toprule
  \textbf{Ideation Stage} & \textbf{Technique} & \textbf{Category} & \textbf{Method} & \textbf{Ref.} \\
  \midrule
  \endfirsthead
  \\
  \\
  \toprule
  \textbf{Ideation Stage} & \textbf{Technique} & \textbf{Category} & \textbf{Method} & \textbf{Ref.} \\
  \midrule
  \endhead
  \multirow{14}{*}{\shortstack{\textbf{Idea} \phantom{xxxxxx} \\ \textbf{Refinement}}} & \multirow{3}{*}{\textbf{Non-LLMs}} & \multirow{2}{*}{\shortstack{\text{Individual \phantom{111 1}} \\ \text{idea refinement}}} & Modification & \cite{xu2024jamplate_G3}, \cite{huang2023causalmapper_G5}, \cite{hou2024c2ideas_D13}, \cite{huang2024plantography_D14}, \cite{yuan2024llmcrit_W15}, \cite{pu2024ideasynth_S1}, \cite{liu2024personaflow_S2} \\
  & & & Elimination & \cite{huang2023causalmapper_G5}, \cite{radensky2024scideator_S3} \\
  \cmidrule{3-5}   
  & & \multirow{1}{*}{\text{Combination}} & Combination & \cite{shaer2024ai_G1}, \cite{suh2024luminate_G13} \\  
  \cmidrule{2-5} 

  & \multirow{11}{*}{\textbf{LLMs}} & \multirow{10}{*}{\shortstack{\text{Individual \phantom{111 1}} \\ \text{idea refinement}}} & Expansion & \cite{di2022idea_G4}, \cite{aikawa2023introducing_G8}, \cite{suh2024luminate_G13}, \cite{kocaballi2023conversational_D1}, \cite{yuan2022wordcraft_W5}, \cite{ghajargar2022redhead_W7}, \cite{pu2024ideasynth_S1}, \cite{cai2023designaid_M1}, \cite{brade2023promptify_M3}, \cite{wang2024lave_M4} \\
  & & & Modification & \cite{di2022idea_G4}, \cite{bhavya2023cam_G16}, \cite{lanzi2023chatgpt_D6}, \cite{ege2024chatgpt_D10}, \cite{junior2023chatgpt_D11}, \cite{goodman2022lampost_W1}, \cite{reza2024abscribe_W13}, \cite{benharrak2024writer_W14}, \cite{lee2024gpt_S6}, \cite{qian2024shape_M5}\\ 
  & & & Substitution & \cite{lanzi2023chatgpt_D6}, \cite{ege2024chatgpt_D10}, \cite{zhang2023visar_W4}, \cite{yuan2022wordcraft_W5}, \cite{reza2024abscribe_W13}, \cite{pu2024ideasynth_S1} \\  
  & & & Rewriting & \cite{di2022idea_G4}, \cite{gonzalez2024collaborative_G7}, \cite{goodman2022lampost_W1}, \cite{yuan2022wordcraft_W5}, \cite{qian2024shape_M5}, \cite{chen2024foundation_M8} \\
  & & & Elimination & \cite{blazevic2024real_G15}, \cite{chung2022talebrush_W3} \\
  & & & Multiple perspectives & \cite{xu2024jamplate_G3}, \cite{aikawa2023introducing_G8}, \cite{suh2024luminate_G13}, \cite{kang2024biospark_G14}, \cite{zhang2023visar_W4}, \cite{benharrak2024writer_W14}, \cite{pu2024ideasynth_S1}, \cite{liu2024personaflow_S2}, \cite{radensky2024scideator_S3}, \cite{wang2023popblends_M2}\\
  & & & Questioning & \cite{xu2024jamplate_G3}, \cite{bilgram2023accelerating_D2}, \cite{wang2023task_D8}, \cite{kim2024towards_W2} \\
  \cmidrule{3-5} 
   
  & & \multirow{1}{*}{\text{Combination}} & Combination & \cite{di2022idea_G4}, \cite{besta2024graph_G12}, \cite{kang2024biospark_G14}, \cite{lanzi2023chatgpt_D6}, \cite{petridis2023anglekindling_W6}, \cite{chen2024foundation_M8}  \\

  \bottomrule
\end{longtable}

As shown in Table \ref{tb:Idea Refinement}, there are two main categories of idea refinement methods: individual idea refinement and idea combination. Individual idea refinement methods include Expansion—extending or continuing existing ideas; Modification—altering existing ideas; Substitution—replacing existing ideas; Rewriting—rephrasing or restating existing ideas; Elimination—removing parts or the entirety of existing ideas; Multiple perspectives—exploring existing ideas from various angles; and Questioning—exploring new ideas by posing questions about existing ones. Idea combination—merging two or more ideas to generate novel concepts.

For individual idea refinement, eight studies utilized non-LLM methods, with the most frequently used method being modification, employed seven times, followed by elimination, used twice. LLMs demonstrated immense potential in this stage, with 32 studies contributing to the idea refinement process through seven methods. The most commonly used idea refinement methods leveraging LLMs were expansion, modification, and multiple perspectives, each utilized in ten studies. Substitution and rewriting were the next most popular, employed in six studies each, followed by questioning, used in four studies, and finally, elimination, which was only utilized in two studies. Regarding idea combination, only two study achieved this through creative team discussions, while six studies employed LLMs to facilitate the process.

The findings highlight the pivotal role of LLMs in the idea refinement stage. By combining a diverse array of methods, LLMs can further refine and enhance the initial ideas generated during the idea generation stage. This enables ideators to explore novel perspectives, identify potential areas for improvement, and ultimately develop ideas that are more innovative and impactful.

\subsubsection{Idea Evaluation}

The idea evaluation stage plays a pivotal role in assessing the quality and potential of generated ideas based on predefined criteria and feedback from diverse perspectives. Among the 61 studies on LLM-assisted ideation activities examined in this review, 46 (75\%) included an idea evaluation stage, underscoring its significance in the overall LLM-assisted ideation process. 

Of the 46 studies that involved an idea evaluation stage, 21 (46\%) employed LLMs to support the evaluation process. Notably, 12 of these studies utilized a combination of human and LLM-based evaluation methods, while only 9 relied solely on LLMs for assessing creative ideas. This finding suggests that, at present, humans prefer to retain decision-making authority in the evaluation of creative ideas rather than delegating it entirely to LLMs.

LLM-based evaluation methods primarily include providing feedback or suggestions on product prototypes \cite{ege2024chatgpt_D10,junior2023chatgpt_D11,chen2024foundation_M8} or text \cite{yuan2024llmcrit_W15,kim2024towards_W2,zhang2023visar_W4,goldi2024intelligent_W10,pu2024ideasynth_S1}; generating virtual personas to offer feedback on academic research ideas \cite{liu2024personaflow_S2}, written content \cite{benharrak2024writer_W14},  or engaging in virtual interviews to provide feedback \cite{kocaballi2023conversational_D1}; integrating LLMs into workflows to facilitate task evaluation \cite{dharuman2023protein_D5,lee2024gpt_S6}; scoring \cite{bhavya2023cam_G16,besta2024graph_G12} and rating ideas \cite{shaer2024ai_G1,aikawa2023introducing_G8} based on specific criteria or functions. Additionally, one study \cite{radensky2024scideator_S3} employed similarity-based reordering of idea-related reference materials to assess the novelty of ideas.

Several studies evaluated ideas using general assessment tools such as strengths and weaknesses \cite{rosenberg2023conversational_G6,kang2024biospark_G14} or specific evaluation criteria, including relevance, innovation, and insightfulness \cite{shaer2024ai_G1}; unique value, advantages, and disadvantages \cite{xu2024jamplate_G3}; originality, efficacy, and feasibility \cite{aikawa2023introducing_G8}; and analogical style, meaningfulness, and novelty \cite{bhavya2023cam_G16}.

It is noteworthy that no studies have yet employed professional evaluation tools such as PPCO or SWOT analysis for assessing ideas generated through LLM-assisted ideation processes. This gap presents an opportunity for future research to explore the potential benefits of integrating these established evaluation frameworks into LLM-assisted ideation activities.

\subsubsection{Multi-idea Evaluation and Selection}

The multi-idea evaluation and selection stage focuses on assessing the large pool of ideas and selecting the most promising ones. Among the 61 studies on LLM-assisted ideation activities examined in this review, 28 (46\%) included a multi-idea evaluation and selection stage, indicating that a substantial proportion of ideation activities generate multiple ideas that are further evaluated and selected to identify the most creative and valuable concepts. 

Among the 28 studies that incorporated the multi-idea evaluation and selection stage, only 4 (14\%) employed LLM-based methods, indicating that people prefer to rely on their own judgment rather than extensively depending on LLMs when selecting creative ideas. This finding presents an opportunity for future research to investigate LLM-based multi-idea evaluation and selection methods.

As shown in Table \ref{tb:Multi-Idea Evaluation and Selection}, numerous non-LLM methods are used for multi-idea evaluation and selection. Comparison is the most commonly employed method, with 19 studies utilizing it, followed by filtering (6 studies), similarity assessment (4 studies), clustering (4 studies), sorting (2 studies), ranking (2 studies), and finally, searching used in one case study. Comparison methods provide multiple ideas for ideators to compare, while filtering and searching methods help ideators identify ideas that meet specific criteria. Similarity assessment and clustering methods structure large numbers of ideas, enabling ideators to evaluate them more clearly. Sorting and ranking methods assist ideators in selecting optimal ideas by ordering them based on specific factors.

\begin{longtable}{>{\raggedright\arraybackslash}p{2.5cm} >
  {\raggedright\arraybackslash}p{3cm} >
  {\raggedright\arraybackslash}p{3cm} >
  {\raggedright\arraybackslash}p{4cm} }
  \caption{Multi-idea Evaluation and Selection Methods} \label{tb:Multi-Idea Evaluation and Selection} \\ 
  \toprule
  \textbf{Ideation Stage} & \textbf{Technique} & \textbf{Method} & \textbf{Ref.} \\
  \midrule
  \endfirsthead
  \\
  \\
  \toprule
  \textbf{Ideation Stage} & \textbf{Technique} & \textbf{Method} & \textbf{Ref.} \\
  \midrule
  \endhead
  \multirow{12}{*}{\shortstack{\textbf{Multi-idea \phantom{xxx}} \\ \textbf{Evaluation \phantom{x x}} \\ \textbf{and Selection}}} & \multirow{9}{*}{\textbf{Non-LLMs}} & Comparison & \cite{shaer2024ai_G1}, \cite{huang2023causalmapper_G5}, \cite{suh2024luminate_G13}, \cite{lanzi2023chatgpt_D6}, \cite{hou2024c2ideas_D13}, \cite{goodman2022lampost_W1}, \cite{chung2022talebrush_W3}, \cite{zhang2023visar_W4}, \cite{yuan2022wordcraft_W5}, \cite{kim2023cells_W9}, \cite{goldi2024intelligent_W10}, \cite{reza2024abscribe_W13}, \cite{benharrak2024writer_W14}, \cite{pu2024ideasynth_S1}, \cite{liu2024personaflow_S2}, \cite{radensky2024scideator_S3}, \cite{cai2023designaid_M1}, \cite{brade2023promptify_M3}, \cite{chen2024foundation_M8}\\
  & & Filtering & \cite{di2022idea_G4}, \cite{huang2023causalmapper_G5}, \cite{suh2024luminate_G13}, \cite{dharuman2023protein_D5}, \cite{radensky2024scideator_S3}, \cite{chen2024foundation_M8} \\ 
  & & Similarity & \cite{bhavya2023cam_G16}, \cite{kim2023cells_W9}, \cite{cai2023designaid_M1}, \cite{wang2023popblends_M2} \\    
  & & Clustering & \cite{gonzalez2024collaborative_G7}, \cite{suh2024luminate_G13}, \cite{brade2023promptify_M3}, \cite{kim2023cells_W9} \\
  & & Sorting & \cite{besta2024graph_G12}, \cite{suh2024luminate_G13} \\ 
  & & Ranking & \cite{besta2024graph_G12}, \cite{wang2023popblends_M2} \\
  & & Searching & \cite{suh2024luminate_G13} \\
  \cmidrule{2-4}  
  & \multirow{3}{*}{\textbf{LLMs}} & Comparison & \cite{xu2024jamplate_G3}, \cite{kang2024biospark_G14} \\
  & & Ranking & \cite{blazevic2024real_G15}, \cite{bhavya2023cam_G16} \\
  & & Rating & \cite{bhavya2023cam_G16} \\
  
  \bottomrule
\end{longtable}

LLMs contribute to the multi-idea evaluation and selection process through three primary methods: comparing multiple ideas \cite{xu2024jamplate_G3,kang2024biospark_G14}, ranking multiple ideas \cite{blazevic2024real_G15,bhavya2023cam_G16}, and rating multiple ideas \cite{bhavya2023cam_G16} to facilitate idea selection.

It is foreseeable that structuring large numbers of ideas by LLMs will be a promising multi-idea evaluation and selection method, particularly in team ideation scenarios. As ideation processes often generate a large number of similar or overlapping ideas, integrating and clustering multiple similar ideas can help streamline the evaluation and selection process, ultimately leading to more effective identification of the most promising concepts.

\subsection{The Iteration of LLM-assisted Ideation Stages}

Iterative processes play a crucial role in ideation, allowing for the continuous refinement and development of ideas. Among the 61 reviewed papers, 28 (46\%) involved iterative ideation activities, highlighting the prevalence and importance of iteration in LLM-assisted ideation.

The majority of these studies, 25 out of 28, demonstrate iterative processes occurring after the evaluation of individual ideas or the selection from multiple ideas. Among them, 13 articles resulted in iteration from the idea evaluation stage, while the other 12 articles iterated forward after experiencing the evaluation and selection of multiple ideas. In these cases, the ideation process returns to the stages of idea generation or idea refinement, enabling users to further explore and enhance their ideas based on the insights gained from the evaluation and selection stages. This iterative loop between idea generation, evaluation, and refinement allows for the progressive development of ideas, leading to more refined and robust concepts.

Three studies \cite{huang2023causalmapper_G5,gonzalez2024collaborative_G7,brade2023promptify_M3} specifically involve selecting and processing individual ideas from idea clusters or multiple ideas. In these cases, the iterative process returns from the multi-idea evaluation and selection stage to the individual idea evaluation stage. This indicates that filtering and evaluating promising ideas at the idea cluster level is a promising approach.

Interestingly, two studies \cite{pu2024ideasynth_S1,liu2024personaflow_S2} focus on the ideation of scientific research ideas. In these cases, idea iteration involves the exploration of new ideas and research starting points from multiple research ideas, representing a return to the scope definition stage of the ideation process. This finding suggests that in the context of scientific research, iterative ideation may involve revisiting and redefining the scope of the problem or research question, allowing for the exploration of novel research directions.

One study \cite{kocaballi2023conversational_D1} employs the creation of different roles to facilitate product ideation. This approach represents an iteration that returns from the idea evaluation stage to the foundational materials collection stage, where the fundamental elements and perspectives that inform the ideation process are established. By iterating on the roles and personas involved in the ideation activity, this study highlights the importance of considering diverse viewpoints and contexts in the development of innovative product ideas.

The varying stages at which iteration is observed in the reviewed literature underscore the dynamic and non-linear nature of ideation processes. While the majority of studies focus on iteration within the core stages of idea generation, refinement, and evaluation, the presence of iteration at earlier stages, such as scope definition and foundational materials collection, emphasizes the need for flexibility and adaptability in LLM-assisted ideation frameworks.

Iterative processes enable ideators to continuously refine and build upon their ideas, incorporating feedback, new insights, and changing perspectives. By accommodating iteration at different stages of the ideation process, the proposed framework facilitates a more comprehensive and dynamic approach to ideation, allowing ideators to explore a broader range of possibilities and arrive at more innovative and well-developed ideas.

As the field of LLM-assisted ideation continues to evolve, it is essential for researchers and designers to recognize the importance of iterative processes and incorporate mechanisms that support iteration at various stages of the ideation journey. By providing ideators with the flexibility to iterate on their ideas, the proposed framework can foster a more engaging, productive, and creative ideation experience.

\newpage
\section{Interaction Design of LLM-assisted Ideation Systems}

\begin{figure}[ht!]
  \centering
  \includegraphics[width=1\textwidth]{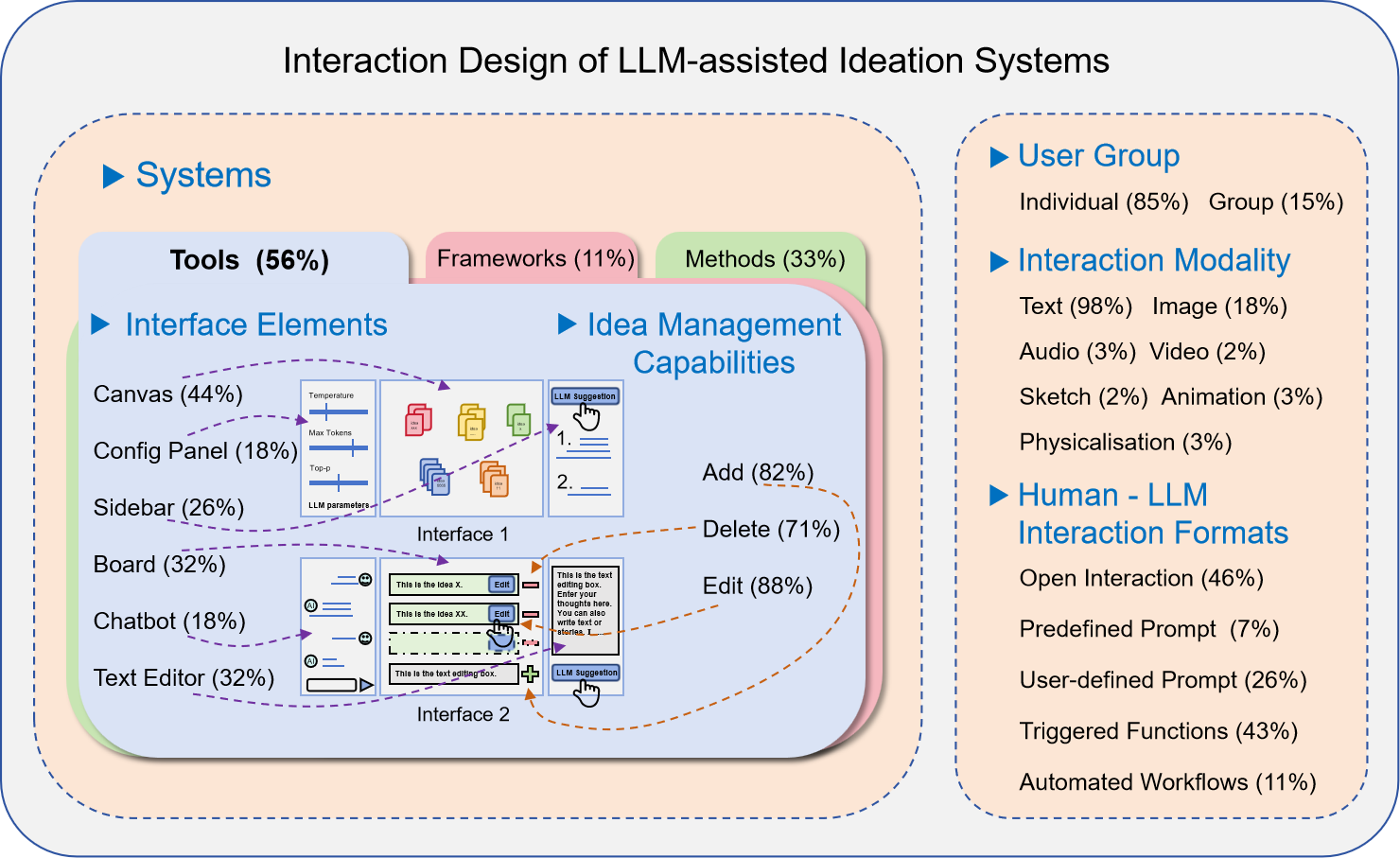}
  \caption{Interaction Design of LLM-assisted Ideation Systems. The systems are categorized into tools (56\%), frameworks (11\%), and methods (33\%). User group analysis reveals a focus on individual ideation support (85\%), while interaction modality is dominated by text (98\%). Interface elements vary, with canvas (44\%), text editor (32\%), and board (32\%) being the most prevalent. Idea management capabilities, such as adding, deleting, and editing ideas, are crucial for user engagement. Human-LLM interaction formats include open interaction (46\%), structured prompts (26\%), triggered functions (43\%), and automated workflows (11\%), each offering distinct trade-offs in user freedom and control.}
  \label{tb:Interaction Framework}
\end{figure}

The rapid advancements in Large Language Models (LLMs) have paved the way for a new generation of ideation systems that harness the power of artificial intelligence to support and enhance human creativity. This paper presents a comprehensive review of LLM-assisted ideation systems (Figure \ref{tb:Interaction Framework}), focusing on their interaction design and the various approaches researchers have taken to facilitate human-AI collaboration in the ideation process.

Our analysis of 61 papers reveals three primary categories of LLM-assisted ideation systems: tools, frameworks, and methods. Tools emphasize the development of user-friendly interfaces, frameworks propose structured approaches to integrate LLMs into ideation workflows, and methods explore novel techniques and strategies for leveraging LLMs in ideation activities. We further examine the interaction design of these systems across five key dimensions: user group, interaction modality, interface elements, idea management capabilities, and human-LLM interaction formats. 

Through this analysis, we aim to provide a comprehensive overview of the current landscape of LLM-assisted ideation systems, identify design patterns and trends, and highlight opportunities for future research and development in this rapidly evolving field.

\newpage
\subsection{Overview of Interaction Design in LLM-assisted Ideation Systems}

Table \ref{tb:Interaction Framework} overviews the interaction design dimensions, with spreadsheet provided separately\footnote[2]{Resources are available at \url{https://doi.org/10.6084/m9.figshare.28440182}}

\begin{table}[H]
  \centering
  \normalsize
  \caption{Interaction Design of LLM-assisted Ideation Systems} 
  \label{tb:Interaction table}
  \resizebox{\textwidth}{!}{
  \begin{tabular}{>
{\raggedright\arraybackslash}m{1cm} >
  {\raggedright\arraybackslash}m{2cm} >
  {\raggedright\arraybackslash}m{1.2cm} >{\raggedright\arraybackslash}m{2cm} >
  {\raggedright\arraybackslash}m{3cm} >  
  {\raggedright\arraybackslash}m{5cm} >{\raggedright\arraybackslash}m{3.3cm} >
  {\raggedright\arraybackslash}m{6.5cm}}
  \toprule
  & & & & \textbf{Interaction} & \textbf{Interface} & \textbf{Idea Management} & \textbf{Human-LLM} \\
  \multirow{1}{*}[0.5\baselineskip]{\textbf{Ref.}} & \multirow{1}{*}[0.5\baselineskip]{\textbf{System}} & \multirow{1}{*}[0.5\baselineskip]{\textbf{Index}} & \multirow{1}{*}[0.5\baselineskip]{\textbf{User Group}} & \textbf{Modality} & \textbf{Element} & \textbf{Capability} & \textbf{Interaction Format} \\
  \midrule

  \cite{heyman2024supermind_G2} & tool & G2 & Individual & Text & board \& sidebar & - & Triggered Functions \\
  \cite{xu2024jamplate_G3} & tool & G3 & Individual & Text & board & Add \& Edit & User-defined Prompt \& Triggered Functions  \\  
  \cite{di2022idea_G4} & tool & G4 & Individual & Text & board & - & Triggered Functions  \\ 
  \cite{huang2023causalmapper_G5} & tool & G5 & Individual & Text & Canvas & Add \& Delete \& Edit & Triggered Functions  \\  
  \cite{gonzalez2024collaborative_G7} & tool & G7 & Group & Text & Canvas \&  Chatbot & Add \& Delete \& Edit & Open Interaction \&  User-defined Prompt  \\  
  \cite{aikawa2023introducing_G8} & tool & G8 & Group & Text \& Audio \& Physicalisation & Augmented Reality & - & Triggered Functions \\ 
  \cite{chen2024asknaturenet_G10} & tool & G10 & Individual & Text & Canvas & - & Triggered Functions  \\
  \cite{suh2024luminate_G13} & tool & G13 & Individual & Text & Text Editor \& Canvas & Add \& Edit & User-defined Prompt \& Triggered Functions  \\
  \cite{kang2024biospark_G14} & tool & G14 & Individual & Text \& Image & board \& sidebar & - & Triggered Functions  \\
  \cite{blazevic2024real_G15} & tool & G15 & Group & Text & board & - & Triggered Functions  \\
  \cite{lee2024generating_D7} & tool & D7 & Individual & Text \& Image & Config Panel \& Chatbot & - &  Open Interaction  \\   
  \cite{lu2024large_D9} & tool & D9 & Individual & Text \& Animation & board \& canvas & - & Automated Workflows  \\ 
  \cite{hou2024c2ideas_D13} & tool & D13 & Individual & Text \& Image & Chatbot \& Config Panel \& Canvas & Edit & Open Interaction \& User-defined Prompt  \\
  \cite{huang2024plantography_D14} & tool & D14 & Individual & Text \& Image & Config Panel \& Canvas & Add \& Delete \& Edit & Triggered Functions  \\
  \cite{goodman2022lampost_W1} & tool & W1 & Individual & Text & Text Editor \& Sidebar & Add \& Delete \& Edit & User-defined Prompt \& Triggered Functions  \\
  \cite{kim2024towards_W2} & tool & W2 & Individual & Text & Text Editor \& Sidebar \& Chatbot & - & Open Interaction \& User-defined Prompt \& Triggered Functions  \\
  \cite{chung2022talebrush_W3} & tool & W3 & Individual & Text \& Sketch & Text Editor \& Canvas & Add \& Delete & Triggered Functions \\  
  \cite{zhang2023visar_W4} & tool & W4 & Individual & Text & Text Editor \& Canvas & Add \& Delete \& Edit & User-defined Prompt \& Triggered Functions  \\ 
  \cite{yuan2022wordcraft_W5} & tool & W5 & Individual & Text & Text Editor \& Sidebar & Add \& Delete \& Edit & User-defined Prompt \& Triggered Functions  \\ 
  \cite{petridis2023anglekindling_W6} & tool & W6 & Individual & Text & board \& sidebar & - & Predefined Prompt   \\ 
  \cite{ghajargar2022redhead_W7} & tool & W7 & Individual & Text & Text Editor \& Canvas & - & Triggered Functions  \\   
  \cite{wan2024felt_W8} & tool & W8 & Individual & Text & Chatbot & - &  Open Interaction  \\ 
  \cite{kim2023cells_W9} & tool & W9 & Individual & Text & Config Panel \& Text Editor & Edit & User-defined Prompt \& Triggered Functions  \\ 
  \cite{goldi2024intelligent_W10} & tool & W10 & Individual & Text & Text Editor \& Sidebar & - & Triggered Functions  \\
  \cite{reza2024abscribe_W13} & tool & W13 & Individual & Text & Text Editor \& Sidebar & Add \& Delete \& Edit & User-defined Prompt \& Triggered Functions  \\
  \cite{benharrak2024writer_W14} & tool & W14 & Individual & Text & Text Editor \& Sidebar & Add \& Delete \& Edit & User-defined Prompt \& Triggered Functions  \\
  \cite{pu2024ideasynth_S1} & tool & S1 & Individual & Text & board \& Canvas & Add \& Delete \& Edit & User-defined Prompt \& Triggered Functions   \\
  \cite{liu2024personaflow_S2} & tool & S2 & Individual & Text & Canvas & Add \& Delete \& Edit & User-defined Prompt \& Triggered Functions  \\
  \cite{radensky2024scideator_S3} & tool & S3 & Individual & Text & board & Add \& Delete & User-defined Prompt \& Triggered Functions  \\  
  \cite{cai2023designaid_M1} & tool & M1 & Individual & Text \& Image & board & - & Triggered Functions  \\
  \cite{wang2023popblends_M2} & tool & M2 & Individual & Text \& Image & board & - & Predefined Prompt  \\
  \cite{brade2023promptify_M3} & tool & M3 & Individual & Text \& Image & Config Panel \& Canvas & - & User-defined Prompt \& Triggered Functions  \\  
  \cite{wang2024lave_M4} & tool & M4 & Individual & Text \& Video & Canvas \& Chatbot & - & Open Interaction \& User-defined Prompt \& Triggered Functions  \\ 
  \cite{qian2024shape_M5} & tool & M5 & Individual & Text \& Animation & Config Panel \& Canvas & Edit & Automated Workflows   \\

  \cite{rosenberg2023conversational_G6} & framework & G6 & Group & Text \& Audio & - & - & Automated Workflows   \\ 
  \cite{besta2024graph_G12} & framework & G12 & Individual & Text & - & - & Automated Workflows  \\
  \cite{bhavya2023cam_G16} & framework & G16 & Individual & Text & - & - & Open Interaction \& Predefined Prompt  \\
  \cite{dharuman2023protein_D5}  & framework & D5 & Individual & Text & - & - & Automated Workflows  \\ 
  \cite{lanzi2023chatgpt_D6}  & framework & D6 & Group & Text & - & - & Automated Workflows   \\ 
  \cite{lee2024gpt_S6}  & framework & S6 & Individual & Text & - & - & Automated Workflows   \\ 
  \cite{chen2024foundation_M8}  & framework & M8 & Individual & Text \& Image & - & - &  Open Interaction  \\

  \cite{shaer2024ai_G1}  & method & G1 & Group & Text & - & - &  Open Interaction  \\
  \cite{schuller2024generating_G9} & method  & G9 & Individual & Text & - & - &  Open Interaction \\ 
  \cite{bernstein2024like_G11}  & method & G11 & Individual & Text & - & - &  Open Interaction  \\
  \cite{paredes2024creative_G17} & method  & G17 & Individual & Audio \& Physicalisation & - & - &  Open Interaction   \\
  \cite{lee2024prompt_G18} & method  & G18 & Individual & Text & - & - &  Open Interaction  \\
  \cite{kocaballi2023conversational_D1}  & method & D1 & Individual & Text & - & - &  Open Interaction   \\
  \cite{bilgram2023accelerating_D2} & method  & D2 & Group & Text & - & - &  Open Interaction  \\
  \cite{sudhakaran2023prompt_D3} & method  & D3 & Individual & Text & - & - &  Open Interaction  \\  
  \cite{todd2023level_D4}  & method & D4 & Individual & Text & - & - &  Open Interaction  \\ 
  \cite{wang2023task_D8} & method  & D8 & Individual & Text & - & - &  Open Interaction \& Predefined Prompt  \\ 
  \cite{ege2024chatgpt_D10} & method  & D10 & Group & Text & - & - &  Open Interaction  \\
  \cite{junior2023chatgpt_D11} & method  & D11 & Group & Text & - & - &  Open Interaction  \\
  \cite{veloso2024forming_D12} & method  & D12 & Individual & Text \& Image & - & - &  Open Interaction  \\
  \cite{li2024mystery_W11} & method  & W11 & Individual & Text & - & - &  Open Interaction \\
  \cite{pividori2024publishing_W12} & method  & W12 & Individual & Text & - & - &  Open Interaction \\
  \cite{yuan2024llmcrit_W15} & method  & W15 & Individual & Text & - & - &  Open Interaction  \\
  \cite{banker2024machine_S4}  & method & S4 & Individual & Text & - & - &  Open Interaction   \\ 
  \cite{chen2024use_S5}  & method & S5 & Individual & Text & - & - &  Open Interaction   \\ 
  \cite{wang2023script_M6} & method  & M6 & Individual & Text \& Image & - & - &  Open Interaction  \\ 
  \cite{wu2023iconshop_M7} & method  & M7 & Individual & Text \& Image & - & - &  Open Interaction  \\ 

  \bottomrule
  \end{tabular}
  }
\end{table}

\subsection{System}

The reviewed literature on LLM-assisted ideation systems can be broadly categorized into three types based on the nature of their contributions: tools, frameworks, and methods. This classification reflects the different approaches researchers have taken to leverage the capabilities of LLMs in supporting ideation processes.

Tools, which account for 34 out of the 61 reviewed papers (56\%), focus on the development of frontend interfaces to facilitate ideation activities. These tools provide users with visual and interactive environments that harness the power of LLMs to generate, refine, and evaluate ideas. By offering intuitive and user-friendly interfaces, these tools aim to enhance the accessibility and effectiveness of LLM-assisted ideation for a wide range of users.

Frameworks, represented by 7 papers (11\%), emphasize the design and implementation of workflows to support ideation processes. Rather than focusing on specific user interfaces, these frameworks propose structured approaches to integrating LLMs into various stages of the ideation pipeline. By establishing clear workflows and defining the roles and interactions of LLMs within these workflows, frameworks provide a systematic and replicable approach to LLM-assisted ideation.

Methods, which encompass 20 papers (33\%), propose novel techniques and strategies for leveraging LLMs in ideation activities without necessarily developing specific interfaces or workflows. These methods focus on the conceptual and theoretical aspects of LLM-assisted ideation, exploring innovative ways to harness the generative and analytical capabilities of LLMs. By proposing new algorithms, prompt engineering techniques, or collaboration strategies, these methods contribute to the advancement of the field and provide a foundation for future tool and framework development.

\subsection{User Group}

According to the Table \ref{tb:Interaction table}, 52 out of the 61 reviewed papers (85\%) concentrate on supporting individual ideation activities, while only 9 papers (15\%) are designed to facilitate group ideation processes. This significant disparity highlights the current focus of LLM-assisted ideation systems on empowering individual users.

The prevalence of individually-oriented systems, such as LaMPost \cite{goodman2022lampost_W1} and LAVE \cite{wang2024lave_M4}, demonstrates the emphasis on leveraging LLMs to enhance personal creativity and idea generation. These systems provide users with powerful tools to independently generate, refine, and evaluate ideas, augmenting their creative capabilities and enabling them to explore a wider range of possibilities.

However, the importance of group ideation cannot be overlooked, particularly in the context of large enterprises and organizations. Collaborative ideation offers several key benefits, including the exchange of diverse perspectives, the fostering of constructive criticism, and the cultivation of a shared sense of ownership over the generated ideas \cite{gonzalez2024collaborative_G7}. By bringing together individuals with different backgrounds, expertise, and viewpoints, team-based ideation can lead to more robust and innovative solutions.

Despite the critical role of group ideation, the interaction table reveals a notable scarcity of LLM-assisted tools specifically designed to support collaborative processes. Among the reviewed papers, only a few systems, such as the Collaborative Canvas \cite{gonzalez2024collaborative_G7} and the Conversational Swarm Intelligence system \cite{rosenberg2023conversational_G6}, are capable of supporting the collaborative processes in group ideation. This also indicates that developing collaborative systems for group ideation still faces unique challenges, which may include developing tools that support synchronous and asynchronous ideation, managing social dynamics, facilitating turn-taking, integrating the contributions of multiple participants, and ensuring equitable participation. There exist significant opportunities for future research and development. By focusing on the design and implementation of systems that cater to the specific needs of collaborative ideation, researchers can fully harness the potential of LLMs in supporting creative processes within teams and organizations.

In conclusion, while the majority of LLM-assisted ideation systems currently prioritize individual ideation support, there is a pressing need for more research and development efforts aimed at facilitating group ideation processes. By addressing the unique challenges of collaborative ideation and leveraging the capabilities of LLMs, future systems can foster more effective and innovative team-based ideation, ultimately driving greater value for enterprises and organizations.

\subsection{Interaction Modality}

The Table \ref{tb:Interaction table} reveals that among the 61 reviewed papers, 43 (70\%) rely on a single mode of interaction, primarily text, for their ideation activities. 17 papers (28\%) employ a combination of text and multimedia modalities, while only 1 paper (2\%) utilizes a non-text-based interaction modality.

The prevalence of text-based interaction in LLM-assisted ideation systems can be attributed to the initial development of LLMs, which focused heavily on text-based applications. This has led to the extensive use of LLMs in domains such as writing and other text-centric tasks. Many of the reviewed papers, such as those dealing with argumentative writing \cite{zhang2023visar_W4} and conceptual involvement \cite{wang2023task_D8}, rely solely on text-based interaction to facilitate ideation processes.

However, in the realm of multimedia creation or domains involving multimedia elements, there is a growing trend towards combining text with various forms of media for interaction. For instance, image generation tasks \cite{brade2023promptify_M3} often employ a combination of text and image-based interaction, while video editing tasks \cite{wang2024lave_M4} incorporate text and video modalities. Group ideation meetings \cite{rosenberg2023conversational_G6} may leverage text and audio interaction to facilitate collaboration. Other interactive modalities explored in the reviewed papers include text and sketch \cite{chung2022talebrush_W3}, text and animation \cite{lu2024large_D9, qian2024shape_M5}, and even text and bodily movements \cite{aikawa2023introducing_G8}.

Interestingly, only one paper in the reviewed set utilizes a non-text-based interaction modality. \cite{paredes2024creative_G17} explores the concept of assisting idea incubation in the context of commuter driving, employing a combination of driver body state and audio interaction. This unique approach highlights the potential for LLM-assisted ideation systems to extend beyond traditional text-based interactions and adapt to specific contextual requirements.

The dominance of text-based interaction in the current landscape of LLM-assisted ideation systems suggests that there is significant room for exploration and innovation in terms of incorporating diverse interaction modalities. By leveraging the capabilities of LLMs in conjunction with multimedia elements and non-textual inputs, future ideation systems can cater to a wider range of creative processes and domains.

Moreover, the integration of multiple interaction modalities can potentially enhance the effectiveness and user experience of LLM-assisted ideation tools. By providing users with the flexibility to interact through various channels, such as text, images, audio, or even physical gestures, these systems can accommodate different preferences, skills, and contexts, ultimately fostering more engaging and productive ideation sessions.

In conclusion, while text-based interaction currently dominates the landscape of LLM-assisted ideation systems, there is a growing recognition of the value of incorporating multimedia and non-textual modalities. As researchers continue to explore and develop innovative interaction modalities, the potential for LLMs to support diverse creative processes across various domains will continue to expand, opening up new opportunities for ideation and collaboration.

\subsection{Interface Element}

Among the 34 LLM-assisted ideation tools reviewed, we identified six primary types of interface components: board, canvas, sidebar, chatbot, config panel, and text editor. These tools are developed using either a single component or a combination of multiple components.

The canvas component is the most widely adopted, with 15 tools (44\%) incorporating it into their design. Canvas interfaces offer the ability to display a rich array of information, including text notes, various forms of diagrams, animations, and more. This versatility allows users to visually organize and manipulate ideas, such as supporting freehand drawing input and other dynamic operations, facilitating a more intuitive and engaging ideation process.

Text editor components are utilized by 11 tools (32\%), enabling users to write and edit text seamlessly. These interfaces provide a familiar environment for users to input, revise, and refine their ideas, making them particularly suitable for writing-centric ideation tasks.

Board components, also present in 11 tools (32\%), serve as the most basic interface element, allowing for the display of text, images, and other forms of information. While relatively simple and lacking the interactive capabilities of canvas interfaces, board interfaces offer a straightforward foundation for presenting and organizing ideas.

Sidebar components are employed by 9 tools (26\%) to extend the main interface and accommodate additional functionalities. Sidebars offer a convenient way to access supplementary features, settings, or information without cluttering the primary workspace, enhancing the overall user experience.

Chatbot components, found in 6 tools (18\%), enable open-ended conversations with LLMs, preserving the flexibility and diversity of interactions. Chatbot interfaces provide a natural and intuitive way for users to engage with the ideation system, fostering a more dynamic and responsive ideation process.

Config panel components, also present in 6 tools (18\%), offer user-friendly parameter configuration options, allowing users to fine-tune the performance of LLMs effectively. By providing accessible controls for adjusting settings, config panels empower users to optimize the ideation system to their specific needs and preferences.

The diversity of interface components employed in LLM-assisted ideation tools reflects the varying requirements and objectives of different ideation processes. By leveraging the strengths of each interface type, these tools can cater to a wide range of user needs, from visual organization and manipulation to text-based input and editing, open-ended conversations, and parameter configuration.

Moreover, the combination of multiple interface components within a single tool can create a more comprehensive and powerful ideation environment. For example, a tool that integrates a canvas, text editor, and sidebar can provide users with a versatile workspace that supports visual ideation, text-based refinement, and easy access to additional features and information.

As the field of LLM-assisted ideation continues to evolve, it is likely that new interface components and combinations will emerge to address the changing needs and preferences of users. Researchers and designers should continue to explore innovative interface solutions that can enhance the effectiveness, usability, and overall experience of ideation tools powered by LLMs.

In conclusion, the interface components employed in LLM-assisted ideation tools play a crucial role in shaping the user experience and facilitating various ideation processes. By understanding the strengths and limitations of each interface type and considering the specific requirements of different ideation tasks, researchers and designers can create more effective and user-friendly tools that harness the power of LLMs to support creative thinking and problem-solving.

\subsection{Idea Management Capability} 

Idea management capabilities are a critical aspect of LLM-assisted ideation tools, as they enable users to interact with and manipulate the ideas generated by the system. Among the 34 tools reviewed, we specifically examined the idea component modules to determine whether they incorporated functionalities for adding, deleting, and editing ideas. These capabilities are crucial in providing users with greater creative freedom and control over the ideation process, facilitating deeper exploration of the creative space.

Our analysis revealed that 17 tools (50\%) developed idea components that considered users' ability to manage ideas. Among these 17 tools, 10 incorporated all three functional components of adding, deleting, and editing ideas \cite{huang2023causalmapper_G5, huang2024plantography_D14, reza2024abscribe_W13}. By offering a comprehensive set of idea management capabilities, these tools empower users to fully engage with the generated ideas, refining and shaping them to align with their creative goals.

Two tools \cite{xu2024jamplate_G3, suh2024luminate_G13} focused on the addition and modification of ideas, while another two tools \cite{chung2022talebrush_W3, radensky2024scideator_S3} prioritized the addition and deletion of ideas. These variations in idea management capabilities reflect the different emphases and objectives of each tool, catering to specific aspects of the ideation process.

Interestingly, three tools \cite{hou2024c2ideas_D13, kim2023cells_W9, qian2024shape_M5} solely considered the modification of ideas. While the ability to edit ideas is undoubtedly valuable, the absence of addition and deletion functionalities may limit users' capacity to fully explore the creative space and manage the generated ideas effectively.

The incorporation of idea management capabilities in LLM-assisted ideation tools is essential for promoting deeper ideation and facilitating the exploration of the creative space. By providing users with the ability to add, delete, and edit ideas, these tools encourage a more active and engaged approach to ideation. Users can build upon the generated ideas, refine them based on their own insights and preferences, and iterate on them to arrive at more sophisticated and well-developed concepts.

Moreover, idea management capabilities contribute to a greater sense of ownership and control over the ideation process. When users can directly manipulate the ideas generated by the system, they are more likely to feel invested in the process and motivated to pursue their creative goals. This increased engagement can lead to more productive ideation sessions and ultimately result in higher-quality ideas.

As the field of LLM-assisted ideation continues to evolve, it is crucial for researchers and designers to recognize the importance of idea management capabilities in shaping the user experience and facilitating effective ideation. By incorporating robust functionalities for adding, deleting, and editing ideas, future tools can provide users with the necessary tools to fully harness the generative power of LLMs and explore the creative space more comprehensively.

In conclusion, idea management capabilities play a vital role in LLM-assisted ideation tools, empowering users to actively engage with and manipulate the ideas generated by the system. The presence of functionalities for adding, deleting, and editing ideas promotes deeper ideation, facilitates the exploration of the creative space, and enhances users' sense of ownership and control over the ideation process. As the development of these tools progresses, prioritizing the incorporation of comprehensive idea management capabilities will be essential for creating more effective and user-centric ideation environments.

\newpage
\subsection{Human-LLM Interaction Format}

The interaction between humans and Large Language Models (LLMs) is a critical aspect of LLM-assisted ideation, serving as the medium through which LLMs support human creativity and determining the degree of freedom users have in their ideation process. Our review of the selected papers reveals four main types of human-LLM interaction formats: open interaction, structured prompt (further divided into predefined prompt and user-defined prompt), triggered functions, and automated workflows.

Open interaction (28 out of 61, or 46\%) is the most flexible and unrestricted format, allowing users to engage in open-ended conversations with LLMs. This interaction format offers the highest degree of freedom, enabling users to explore and express their ideas without predefined constraints. Among the reviewed papers, 22 out of 29 framework and method-type ideation activities employ open interaction, as they do not involve the development of a specific tool interface. In these cases, users communicate with LLMs through a dialogue interface to facilitate their ideation process. Additionally, 6 tool-type ideation activities incorporate open interaction alongside their developed interfaces, enriching the diversity of human-LLM interaction, as reflected in the chatbot interface component discussed in the interface elements section.

Structured prompt interaction involves users interacting with LLMs through structured prompts, which can be further classified into predefined prompts and user-defined prompts. Predefined prompts are fixed and unchangeable, while user-defined prompts allow users to customize all or part of the prompt content.

Predefined prompts are used in 4 out of 61 reviewed papers (7\%). Two tool-type studies \cite{petridis2023anglekindling_W6, wang2023popblends_M2} incorporate predefined prompts into their backend systems. The other two studies \cite{bhavya2023cam_G16, wang2023task_D8} involve users directly interacting with LLMs using structured, predefined prompts.

User-defined (16 out of 61, or 26\%) prompts are exclusively found in tool-type ideation activities, with 16 out of 34 tool-type studies employing this interaction format. These tools provide users with semi-structured or blank prompt editing areas, offering a higher degree of freedom in human-LLM interaction compared to predefined prompts.

Triggered functions (26 out of 61, or 43\%) are also unique to tool-type ideation activities, with 26 out of 34 tool-type studies utilizing this interaction format. In triggered functions, prompts or instructions for LLMs are integrated into buttons or other interface elements, enabling the execution of specific functions without requiring users to directly engage in text-based interaction with LLMs.

Automated workflows are employed in 7 out of 61 reviewed papers (11\%), with 5 of these studies being framework-type ideation activities. In these cases, LLMs are integrated into workflows or task flows, allowing users to engage in ideation activities without direct interaction with LLMs. Instead, LLMs automatically interact with other components within the workflow.

The diverse range of human-LLM interaction formats observed in the reviewed papers highlights the various approaches researchers and designers have taken to facilitate LLM-assisted ideation. Each interaction format offers distinct advantages and trade-offs in terms of user freedom, control, and ease of use.

Open interaction provides the greatest flexibility and freedom, allowing users to explore their ideas without constraints. However, this format may require users to have a deeper understanding of LLMs and their capabilities to effectively leverage them for ideation.

Structured prompts, both predefined and user-defined, offer a more guided approach to ideation, helping users focus their interactions with LLMs. While predefined prompts provide a straightforward and accessible way to engage with LLMs, they may limit users' ability to fully express their creative intent. User-defined prompts strike a balance between structure and freedom, enabling users to customize their interactions while still benefiting from the guidance provided by the prompts.

Triggered functions and automated workflows represent more streamlined approaches to LLM-assisted ideation, abstracting away the complexity of direct interaction with LLMs. These formats can lower the barrier to entry for users and provide a more efficient and targeted ideation experience. However, they may also reduce users' sense of control and ownership over the ideation process.

As the field of LLM-assisted ideation continues to evolve, it is crucial for researchers and designers to carefully consider the implications of different human-LLM interaction formats on user experience, creativity, and outcomes. By understanding the strengths and limitations of each format and aligning them with the specific goals and requirements of ideation activities, future systems can be designed to effectively harness the power of LLMs while empowering users to fully express their creativity.

In conclusion, the diverse human-LLM interaction formats employed in LLM-assisted ideation systems reflect the ongoing efforts to create effective and engaging user experiences. Each interaction format presents unique advantages and challenges, influencing the degree of user control, flexibility, and accessibility. By aligning the choice of interaction format with the specific goals and requirements of ideation activities, future systems can optimize the collaboration between humans and LLMs, ultimately fostering more creative and productive ideation experiences.

\newpage
\section{User Study of LLM-assisted Ideation Systems}

\begin{figure}[ht!]
  \centering
  \includegraphics[width=1\textwidth]{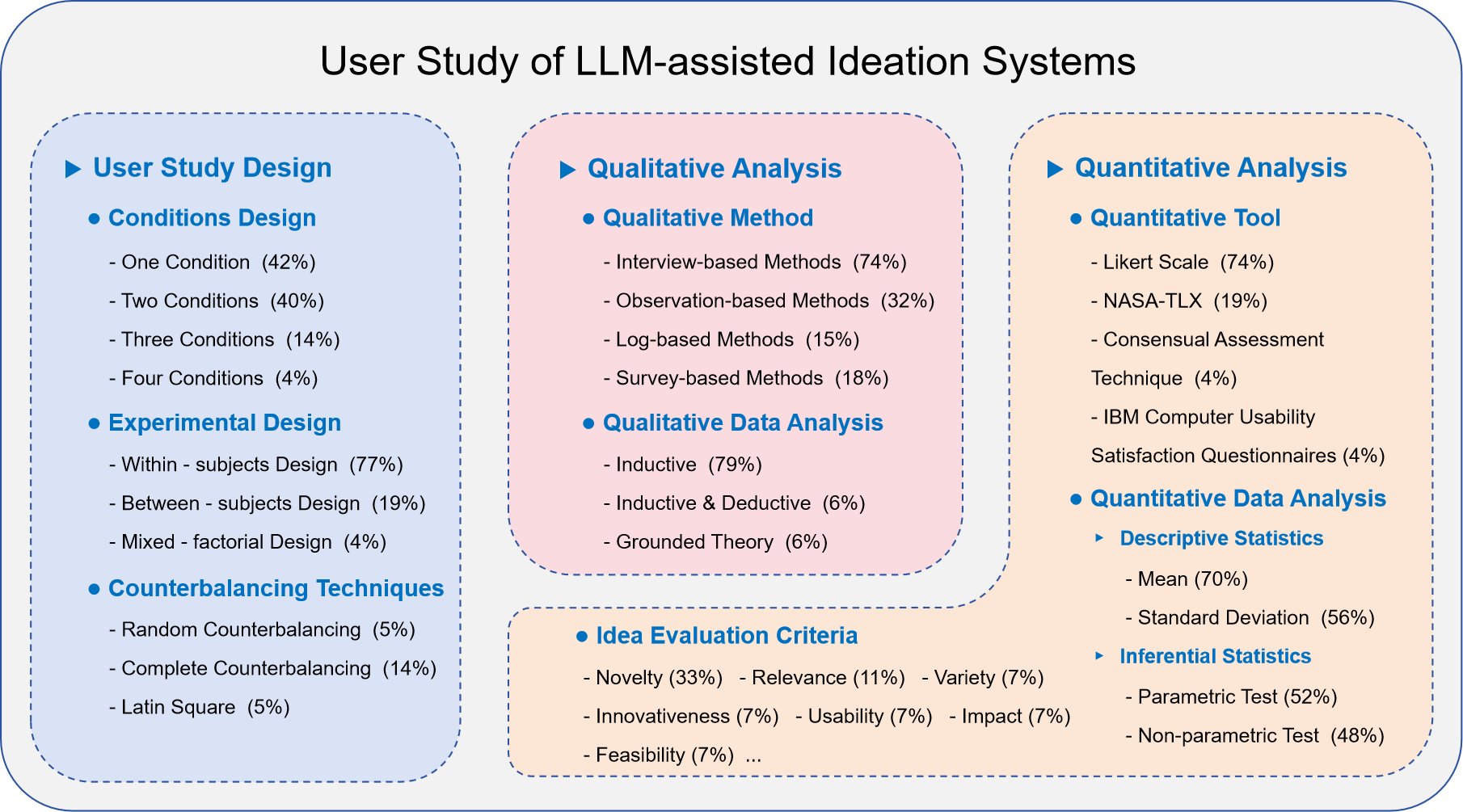}
  \caption{User Study of LLM-assisted Ideation Systems. The analysis focuses on three key aspects: user study design, qualitative analysis, and quantitative analysis. User study design examines the number of conditions investigated and the experimental design approaches adopted. Qualitative analysis methods include interview-based methods (58\%), observation-based methods (30\%), and log-based methods (12\%), with inductive approaches being the primary qualitative data analysis techniques. Quantitative analysis utilizes tools such as Likert scales (58\%), NASA-TLX (30\%), and Consensual Assessment Technique (12\%). Descriptive statistics (44\% mean, 35\% standard deviation) and inferential statistics (33\% parametric tests, 30\% non-parametric tests) are used to analyse the quantitative data, with novelty, relevance, and functionality being the most frequently used evaluation criteria for ideas.}
  \label{tb:User Study Framework}
\end{figure}

Among the 61 papers reviewed, 43 (70\%) conducted a user study, highlighting the vital role of user studies in evaluating the effectiveness, usability, and user experience of LLM-assisted ideation systems. To gain a comprehensive understanding of how these systems are assessed and perceived by users, we conducted a systematic review of the user study methodologies employed in the selected papers (Figure \ref{tb:User Study Framework}). Our analysis focused on three key aspects: user study design, qualitative analysis, and quantitative analysis.

By synthesizing the user study methodologies employed in the reviewed papers, we aim to provide a comprehensive overview of the current practices in evaluating LLM-assisted ideation systems from a user perspective. This analysis highlights the diverse approaches taken by researchers to assess the effectiveness and user reception of these systems, as well as the key dimensions considered in their evaluation. The insights gained from this review can inform the design and execution of future user studies in this domain, ultimately contributing to the development of more user-centric and effective LLM-assisted ideation systems.

\subsection{User Study Design}

The design of user experiments plays a crucial role in ensuring the validity and reliability of the findings. As shown in Table \ref{tb:Experiment Design}, among the 43 papers that conducted user study, 18 studies (42\%) focused on a single condition, while 17 studies (40\%) investigated two conditions. Additionally, 6 studies (14\%) explored three conditions, and 2 studies (4\%) examined four conditions.

The choice between single-condition and multi-condition experiments reflects the researchers' objectives and the complexity of the investigated phenomena. Single-condition studies typically involve the evaluation of a specific LLM-assisted ideation tool, method, or framework developed within the study. These experiments aim to provide an in-depth exploration of the proposed approach, assessing its performance, user reception, and potential impact on the ideation process. By focusing on a single condition, researchers can gather detailed insights into the strengths, weaknesses, and user experiences associated with their proposed solution.

\begin{table}[H]
  \centering
  \small
  \caption{User Study Design}
  \label{tb:Experiment Design}
  \resizebox{\textwidth}{!}{
  \begin{tabular}{>{\raggedright\arraybackslash}m{4.5cm} >{\raggedright\arraybackslash}m{4.5cm} >
  {\raggedright\arraybackslash}m{4.5cm} >
  {\raggedright\arraybackslash}m{3cm}}
  \toprule
  \textbf{Conditions Design} & \textbf{Experimental Design} & \textbf{Counterbalancing Technique} & \textbf{Ref.} \\
  \midrule
  One Condition & - & - & \cite{shaer2024ai_G1}, \cite{xu2024jamplate_G3}, \cite{rosenberg2023conversational_G6}, \cite{bernstein2024like_G11}, \cite{suh2024luminate_G13}, \cite{kang2024biospark_G14}, \cite{blazevic2024real_G15}, \cite{bhavya2023cam_G16}, \cite{kocaballi2023conversational_D1}, \cite{lanzi2023chatgpt_D6}, \cite{hou2024c2ideas_D13}, \cite{kim2024towards_W2}, \cite{chung2022talebrush_W3}, \cite{ghajargar2022redhead_W7}, \cite{wan2024felt_W8}, \cite{liu2024personaflow_S2}, \cite{wang2024lave_M4}, \cite{qian2024shape_M5} \\
  \midrule
  & mixed-factorial design & Latin Square & \cite{cai2023designaid_M1} \\
  \cmidrule{2-4}
  & between-subjects design & - & \cite{ege2024chatgpt_D10}, \cite{goodman2022lampost_W1}, \cite{kim2023cells_W9}, \cite{benharrak2024writer_W14} \\
  \cmidrule{2-4}
  Two Conditions & & - & \cite{xu2024jamplate_G3}, \cite{schuller2024generating_G9}, \cite{lee2024prompt_G18}, \cite{radensky2024scideator_S3}, \cite{banker2024machine_S4}, \cite{brade2023promptify_M3} \\
  \cmidrule{3-4}
  & \multirow{2}{*}[1.2\baselineskip]{within-subjects design} & random counterbalancing & \cite{huang2024plantography_D14} \\
  \cmidrule{3-4}
  & & complete counterbalancing & \cite{paredes2024creative_G17}, \cite{reza2024abscribe_W13}, \cite{petridis2023anglekindling_W6}, \cite{pu2024ideasynth_S1}, \cite{wang2023popblends_M2} \\  
  \midrule
  & mixed-factorial design & complete counterbalancing & \cite{heyman2024supermind_G2} \\
  \cmidrule{2-4}
  & between-subjects design & - & \cite{chen2024asknaturenet_G10}, \cite{wang2023task_D8}, \cite{chen2024foundation_M8} \\
  \cmidrule{2-4}
  \multirow{2}{*}[1.2\baselineskip]{Three Conditions} & & Latin Square & \cite{zhang2023visar_W4}  \\
  \cmidrule{3-4}
  & \multirow{2}{*}[1.2\baselineskip]{within-subjects design} & random counterbalancing & \cite{yuan2022wordcraft_W5}  \\
  \midrule  
  & between-subjects design & - & \cite{goldi2024intelligent_W10} \\
  \cmidrule{2-4}
  \multirow{2}{*}[1.2\baselineskip]{Four Conditions} & within-subjects design & - & \cite{wu2023iconshop_M7} \\
  \bottomrule
  \end{tabular}
  }
\end{table}

On the other hand, multi-condition experiments are designed to enable comparative analyses between different approaches or variations of an LLM-assisted ideation system. These experiments may involve comparing the developed tool or method against a baseline or existing solution, allowing researchers to assess the relative effectiveness and user preferences between the two conditions. Additionally, multi-condition experiments may investigate the impact of different configurations or modules within the proposed system. For example, researchers may compare the performance of their fully-featured tool against a version with only a subset of its modules or functionalities, shedding light on the contribution of specific components to the overall ideation process.

Within the multi-condition experiments, 15 papers (77\%) employed a within-subjects design, where each participant experienced all the conditions being investigated. This design allows for direct comparisons of user experiences and reduces the impact of individual differences on the results. Conversely, 8 papers (19\%) utilized a between-subjects design, assigning each participant to only one of the conditions. This approach minimizes potential carry-over effects and learning biases that may arise from exposure to multiple conditions. Only 2 papers (4\%) adopted a mixed-factorial design, combining elements of both within-subjects and between-subjects designs to investigate the interactions between different factors.

To control for potential order effects in within-subjects and mixed-factorial designs, researchers employed various counterbalancing techniques. Two papers (5\%) used random counterbalancing, where the order of conditions was randomized for each participant. This approach helps to distribute any order effects evenly across the conditions. Six papers (14\%) utilized complete counterbalancing designs, ensuring that all possible order permutations were equally represented in the study. Two paper (5\%) employed a Latin square design, a systematic method for assigning participants to different condition sequences.

The choice of experiment design and counterbalancing techniques reflects the researchers' efforts to minimize confounding factors and enhance the internal validity of their studies. By carefully considering these methodological aspects, the reviewed papers aim to provide robust and reliable findings that contribute to the understanding of LLM-assisted ideation from a user perspective.

\subsection{Qualitative Analysis}

Among the 43 papers that conducted user studies, 34 (79\%) employed qualitative analysis methods to gain in-depth insights into user experiences, perceptions, and behaviors related to LLM-assisted ideation. As shown in Table \ref{tb:Qualitative analysis}, the qualitative methods used in these studies can be broadly categorized into five types: interview-based methods, observation-based methods, log-based methods, survey-based methods, and other methods.

\begin{table}[H]
  \centering
  \small
  \caption{Qualitative Analysis}
  \label{tb:Qualitative analysis}
  \resizebox{\textwidth}{!}{
  \begin{tabular}{>{\raggedright\arraybackslash}m{4.5cm} >{\raggedright\arraybackslash}m{4.5cm} >
  {\raggedright\arraybackslash}m{4.5cm} >
  {\raggedright\arraybackslash}m{3cm}}
  \toprule
  \textbf{Qualitative Method} & \textbf{Qualitative Form} & \textbf{Qualitative Data Analysis} & \textbf{Ref.} \\
  \midrule
  & & inductive & \cite{xu2024jamplate_G3}, \cite{suh2024luminate_G13}, \cite{kang2024biospark_G14}, \cite{blazevic2024real_G15}, \cite{wang2023task_D8}, \cite{huang2024plantography_D14}, \cite{chung2022talebrush_W3}, \cite{zhang2023visar_W4}, \cite{yuan2022wordcraft_W5}, \cite{petridis2023anglekindling_W6}, \cite{kim2023cells_W9}, \cite{benharrak2024writer_W14}, \cite{pu2024ideasynth_S1}, \cite{liu2024personaflow_S2}, \cite{radensky2024scideator_S3}, \cite{wang2023popblends_M2}, \cite{brade2023promptify_M3}, \cite{wang2024lave_M4}, \cite{qian2024shape_M5}, \cite{chen2024foundation_M8}\\
  \cmidrule{3-4}
  \multirow{2}{*}[1.2\baselineskip]{Interview-based Methods} & \multirow{2}{*}[2.2\baselineskip]{semi-structured interview} & inductive \& deductive & \cite{goodman2022lampost_W1}, \cite{reza2024abscribe_W13} \\
  \cmidrule{3-4}
  & & grounded theory & \cite{wan2024felt_W8} \\
  \cmidrule{2-4}  
  & unstructured Interview & inductive & \cite{huang2023causalmapper_G5} \\
  \cmidrule{2-4} 
  & simulated Interview & inductive & \cite{kocaballi2023conversational_D1} \\
  \midrule  
  & & inductive & \cite{blazevic2024real_G15},  \cite{brade2023promptify_M3} \\
  \cmidrule{3-4}  
  & \multirow{2}{*}[1.2\baselineskip]{observation} & inductive \& deductive & \cite{reza2024abscribe_W13} \\
  \cmidrule{2-4}
  Observation-based Methods & think-aloud & inductive & \cite{kang2024biospark_G14}, \cite{hou2024c2ideas_D13}, \cite{chung2022talebrush_W3}, \cite{benharrak2024writer_W14}, \cite{liu2024personaflow_S2}, \cite{wang2024lave_M4} \\
  \cmidrule{2-4}
  & & inductive & \cite{kang2024biospark_G14}, \cite{kim2024towards_W2}, \cite{chung2022talebrush_W3} \\
  \cmidrule{3-4} 
  & \multirow{2}{*}[1.2\baselineskip]{screen recordings} & grounded theory & \cite{wan2024felt_W8} \\  
  \midrule  
  & system log & inductive & \cite{huang2023causalmapper_G5}, \cite{pu2024ideasynth_S1}, \cite{liu2024personaflow_S2} \\
  \cmidrule{2-4}    
  Log-based Methods & prompts log & inductive & \cite{shaer2024ai_G1} \\  
  \cmidrule{2-4}  
  & chat log & grounded theory & \cite{ege2024chatgpt_D10} \\
  \midrule 
  & & - & \cite{lanzi2023chatgpt_D6} \\
  \cmidrule{3-4}   
  \multirow{2}{*}[1.2\baselineskip]{Survey-based Methods} & \multirow{2}{*}[1.2\baselineskip]{survey} & inductive & \cite{shaer2024ai_G1}, \cite{xu2024jamplate_G3}, \cite{schuller2024generating_G9}, \cite{bernstein2024like_G11}, \cite{suh2024luminate_G13} \\
  \midrule
  & autoethnography & inductive & \cite{ghajargar2022redhead_W7} \\  
  \cmidrule{2-4} 
  \multirow{2}{*}[1.2\baselineskip]{Other Methods} & case study method & - & \cite{lee2024prompt_G18} \\  

  \bottomrule
  \end{tabular}
  }
\end{table}

Interview-based methods (25 out of 34, or 74\%) were the most commonly employed, with semi-structured interviews being the predominant form. Semi-structured interviews allow researchers to explore specific topics of interest while providing flexibility for participants to share their thoughts and experiences. For example, two studies \cite{xu2024jamplate_G3, suh2024luminate_G13} utilized semi-structured interviews to gather user feedback on their proposed LLM-assisted ideation tools. Other forms of interviews, such as unstructured interviews \cite{huang2023causalmapper_G5} and simulated interviews \cite{kocaballi2023conversational_D1}, were also used to capture more open-ended and naturalistic user responses.

Observation-based methods (11 out of 34, or 32\%) were employed to study user behavior and interactions with LLM-assisted ideation systems in real-time. Techniques such as direct observation \cite{blazevic2024real_G15}, think-aloud protocols \cite{kang2024biospark_G14, hou2024c2ideas_D13}, and screen recordings \cite{kim2024towards_W2} provided researchers with valuable insights into the usability, user experience, and cognitive processes involved in ideation tasks. These methods allowed researchers to identify patterns, challenges, and opportunities for improvement in the design and implementation of LLM-assisted ideation tools.

Log-based methods (5 out of 34, or 15\%) involved the analysis of data generated by the LLM-assisted ideation systems themselves. System logs \cite{huang2023causalmapper_G5, pu2024ideasynth_S1}, prompts logs \cite{shaer2024ai_G1}, and chat logs \cite{ege2024chatgpt_D10} provided researchers with objective data on user interactions, input patterns, and system responses. By examining these logs, researchers could gain insights into user behavior, preferences, and the effectiveness of different system components in supporting the ideation process.

Survey-based methods (6 out of 34, or 15\%), such as questionnaires and surveys, were used to collect user opinions, ratings, and feedback on various aspects of LLM-assisted ideation systems. For example, studies \cite{lanzi2023chatgpt_D6, bernstein2024like_G11} employed surveys to assess user satisfaction, perceived usefulness, and overall experience with the proposed tools and methods.

Other methods (2 out of 34, or 6\%), including autoethnography \cite{ghajargar2022redhead_W7} and case study methods \cite{lee2024prompt_G18}, were also used to provide rich, contextualized accounts of user experiences and to explore the application of LLM-assisted ideation in specific domains or use cases.

Regarding qualitative data analysis techniques, the majority of the studies (27 out of 34, or 79\%) employed inductive approaches, such as thematic analysis or content analysis. Inductive analysis involves identifying themes, patterns, and categories that emerge from the data itself, allowing researchers to develop a deep understanding of user experiences and perspectives. Two studies \cite{goodman2022lampost_W1, reza2024abscribe_W13} combined inductive and deductive analysis, using predefined categories or frameworks to guide the analysis while remaining open to new insights. Additionally, two studies \cite{wan2024felt_W8, ege2024chatgpt_D10} utilized grounded theory, a systematic methodology for developing theories grounded in the data.

\subsection{Quantitative Analysis}

Among the 43 papers that conducted user studies, 27 (63\%) employed quantitative analysis methods to assess various aspects of LLM-assisted ideation systems. 

\begin{table}[H]
  \centering
  \small
  \caption{Quantitative Analysis}
  \label{tb:Quantitative analysis}
  \resizebox{\textwidth}{!}{
  \begin{tabular}{>{\raggedright\arraybackslash}m{4.5cm} >{\raggedright\arraybackslash}m{4.5cm} >
  {\raggedright\arraybackslash}m{4.5cm} >
  {\raggedright\arraybackslash}m{3cm}}
  \toprule
  \textbf{Quantitative Tool} & \textbf{Rating Scale} & \textbf{Descriptive Statistics} & \textbf{Ref.} \\
  \midrule
  & 4-point Likert scale & mean & \cite{bhavya2023cam_G16} \\
  \cmidrule{2-4}
  & & - & \cite{heyman2024supermind_G2}, \cite{xu2024jamplate_G3}, \cite{huang2024plantography_D14}, \cite{liu2024personaflow_S2} \\
  \cmidrule{3-4}
  & & mean & \cite{schuller2024generating_G9}, \cite{yuan2022wordcraft_W5} \\
  \cmidrule{3-4}
  \multirow{2}{*}[1\baselineskip]{Likert Scale} & \multirow{2}{*}[2\baselineskip]{5-point Likert scale} & mean \& SD & \cite{shaer2024ai_G1}, \cite{suh2024luminate_G13}, \cite{wang2023task_D8}, \cite{hou2024c2ideas_D13}, \cite{benharrak2024writer_W14},  \cite{banker2024machine_S4}, \cite{brade2023promptify_M3} \\ 
  \cmidrule{2-4}
  &  &  mean & \cite{zhang2023visar_W4} \\
  \cmidrule{3-4}  
  & \multirow{2}{*}[1.5\baselineskip]{7-point Likert scale} & mean \& SD & \cite{goodman2022lampost_W1}, \cite{petridis2023anglekindling_W6}, \cite{kim2023cells_W9}, \cite{radensky2024scideator_S3}, \cite{banker2024machine_S4}, \cite{cai2023designaid_M1}, \cite{wang2024lave_M4}\\
  \midrule  
  & absolute rating scale & - & \cite{chen2024foundation_M8} \\
  \cmidrule{2-4}  
  NASA-TLX & & - & \cite{reza2024abscribe_W13}, \cite{pu2024ideasynth_S1} \\
  \cmidrule{3-4}
  & \multirow{2}{*}[1.2\baselineskip]{7-point Likert scale} & mean \& SD & \cite{wang2023popblends_M2}, \cite{brade2023promptify_M3} \\
  \midrule  
  Consensual Assessment Technique & 5-point Likert scale & - & \cite{chen2024asknaturenet_G10} \\
  \midrule
  IBM Computer Usability Satisfaction Questionnaires & 3-point Likert scale & - & \cite{blazevic2024real_G15} \\

  \bottomrule
  \end{tabular}
  }
\end{table}

As shown in Table \ref{tb:Quantitative analysis}, the quantitative tools used in these studies can be broadly categorized into four types: Likert Scale, NASA-TLX, Consensual Assessment Technique, and IBM Computer Usability Satisfaction Questionnaires.

Likert Scale (20 out of 27, or 74\%) was the most commonly used quantitative tool, with variations in the number of points on the scale. Studies employed 4-point \cite{bhavya2023cam_G16}, 5-point \cite{shaer2024ai_G1, suh2024luminate_G13}, and 7-point \cite{goodman2022lampost_W1, petridis2023anglekindling_W6} Likert scales to measure user perceptions, attitudes, and evaluations of different aspects of the LLM-assisted ideation systems. The choice of scale granularity reflects the researchers' intended level of precision in capturing user responses.

NASA-TLX (Task Load Index) (5 out of 27, or 19\%) was used in five studies \cite{chen2024foundation_M8, reza2024abscribe_W13, pu2024ideasynth_S1, wang2023popblends_M2, brade2023promptify_M3} to assess the subjective workload experienced by users while interacting with the ideation systems. This tool employs both an absolute rating scale and a 7-point Likert scale to measure various dimensions of workload, such as mental demand, physical demand, and effort.

The Consensual Assessment Technique (CAT) (1 out of 27, or 4\%) was used in one study \cite{chen2024asknaturenet_G10} to evaluate the creativity of ideas generated with the assistance of an LLM-based system. CAT involves multiple expert judges independently rating the creativity of ideas using a 5-point Likert scale, providing a reliable and validated measure of creativity.

IBM Computer Usability Satisfaction Questionnaires (1 out of 27, or 4\%) were employed in one study \cite{blazevic2024real_G15} to assess user satisfaction with an LLM-assisted ideation system. This tool uses a 3-point Likert scale to measure various aspects of usability, such as ease of use, learnability, and overall satisfaction.

\begin{table}[H]
  \centering
  \small
  \caption{Inferential Statistics}
  \label{tb:Inferential Statistics}
  \resizebox{\textwidth}{!}{
  \begin{tabular}{>{\raggedright\arraybackslash}m{4.5cm} >{\raggedright\arraybackslash}m{4.5cm} >
  {\raggedright\arraybackslash}m{4.5cm} >
  {\raggedright\arraybackslash}m{3cm}}
  \toprule
  \textbf{Category} & \textbf{Inferential Statistics} & \textbf{Method} & \textbf{Ref.} \\
  \midrule
  & Normality Test & Shapiro-Wilk Test & \cite{shaer2024ai_G1}, \cite{kim2023cells_W9} \\
  \cmidrule{2-4}
  & & Independent T-test & \cite{kim2023cells_W9}, \cite{cai2023designaid_M1} \\
  \cmidrule{3-4}
  & \multirow{2}{*}[0.6\baselineskip]{Mean Comparison Test} & Paired-sample T-test & \cite{yuan2022wordcraft_W5}, \cite{pu2024ideasynth_S1}, \cite{liu2024personaflow_S2}, \cite{radensky2024scideator_S3}, \cite{cai2023designaid_M1}  \\
  \cmidrule{3-4}
  \multirow{2}{*}[1\baselineskip]{Parametric Test} &  & One-tailed T-test & \cite{reza2024abscribe_W13} \\ 
  \cmidrule{2-4}
  &  & Welch’s ANOVA & \cite{heyman2024supermind_G2} \\
  \cmidrule{3-4}  
  & Analysis of Variance & ANOVA & \cite{chen2024asknaturenet_G10}, \cite{zhang2023visar_W4}, \cite{goldi2024intelligent_W10}, \cite{wu2023iconshop_M7} \\
  \cmidrule{3-4}
  & & MANOVA & \cite{pu2024ideasynth_S1} \\
  \cmidrule{2-4}  
  & Correlation Test & Pearson Correlation & \cite{shaer2024ai_G1}, \cite{liu2024personaflow_S2}, \cite{wang2023popblends_M2} \\
  \midrule
  & & Wilcoxon Rank-Sum Test / Mann-Whitney U Test & \cite{schuller2024generating_G9}, \cite{goodman2022lampost_W1}, \cite{yuan2022wordcraft_W5}, \cite{kim2023cells_W9}, \cite{goldi2024intelligent_W10} \\
  \cmidrule{3-4}  
  & \multirow{2}{*}[1.5\baselineskip]{Rank Test} & Wilcoxon Signed-Rank Test & \cite{pu2024ideasynth_S1}, \cite{radensky2024scideator_S3}, \cite{banker2024machine_S4}, \cite{wang2023popblends_M2}, \cite{brade2023promptify_M3} \\
  \cmidrule{2-4} 
  \multirow{2}{*}[2\baselineskip]{Non-parametric Test} & & Kruskal-Wallis Test & \cite{chen2024asknaturenet_G10} \\
  \cmidrule{3-4} 
  & \multirow{2}{*}[1.2\baselineskip]{Multiple Group Comparison} & Friedman’s Test & \cite{hou2024c2ideas_D13} \\
  \cmidrule{2-4}   
  & Post-hoc Test & Games-Howell Test & \cite{heyman2024supermind_G2} \\
  \midrule  

  \bottomrule
  \end{tabular}
  }
\end{table}

Quantitative data analysis methods can be divided into two main categories: descriptive statistics and inferential statistics. As shown in the Table \ref{tb:Quantitative analysis}, 19 studies (70\%) used descriptive statistics to summarize and present their quantitative data. Of these, 4 studies used only mean values, while 15 studies used a combination of mean and standard deviation to provide a more comprehensive description of the data.

Inferential statistical methods were employed in 19 studies (70\%) to draw conclusions and test hypotheses based on the quantitative data. As shown in Table \ref{tb:Inferential Statistics}, these methods can be further classified into parametric tests and non-parametric tests. Parametric tests (14 out of 27, or 52\%), which assume a normal distribution of the data, include Normality Tests (e.g., Shapiro-Wilk Test), Mean Comparison Tests (e.g., Independent T-test, Paired-sample T-test, One-tailed T-test), Analysis of Variance (e.g., Welch's ANOVA, ANOVA, MANOVA), and Correlation Tests (e.g., Pearson Correlation). Non-parametric tests (13 out of 27, or 48\%), which do not assume a normal distribution of the data, include Rank Tests (e.g., Wilcoxon Rank-Sum Test or Mann-Whitney U Test, Wilcoxon Signed-Rank Test), Multiple Group Comparison (e.g., Kruskal-Wallis Test, Friedman's Test), and Post-hoc Tests (e.g., Games-Howell Test).

Regarding the evaluation criteria used in quantitative analyses of ideas, the Table \ref{tb:Criteria} reveals that novelty was the most frequently used criterion, appearing in 6 studies. Other related terms, such as innovativeness and uniqueness, were used in 3 studies. Relevance was the second most common criterion, used in 3 studies, while variety was used in 2 studies.

\begin{table}[H]
  \centering
  \small
  \caption{Evaluation Criteria}
  \label{tb:Criteria}
  \resizebox{\textwidth}{!}{
  \begin{tabular}{>{\raggedright\arraybackslash}m{4cm} >{\raggedright\arraybackslash}m{2.5cm} >
  {\raggedright\arraybackslash}m{7.5cm} >
  {\raggedright\arraybackslash}m{2.5cm}}
  \toprule
  \textbf{Category} & \textbf{Criterion} & \textbf{Description} & \textbf{Ref.} \\
  \midrule
  Novelty & Novelty & Evaluates whether the idea demonstrates originality and offers a fresh perspective or concept. & \cite{chen2024asknaturenet_G10}, \cite{bhavya2023cam_G16}, \cite{wang2023task_D8}, \cite{pu2024ideasynth_S1}, \cite{radensky2024scideator_S3}, \cite{wu2023iconshop_M7} \\
  \cmidrule{2-4}
  & Innovativeness & Assesses the degree to which the idea introduces innovative approaches that transcend existing norms. & \cite{shaer2024ai_G1}, \cite{heyman2024supermind_G2} \\
  \cmidrule{2-4} 
  & Uniqueness & Examines whether the idea stands out distinctly from other concepts or generated outputs. & \cite{wu2023iconshop_M7}  \\
  \midrule 
  Relevance & Relevance & Measures the alignment of the idea with the given task, topic, or research objectives. & \cite{shaer2024ai_G1}, \cite{pu2024ideasynth_S1}, \cite{banker2024machine_S4} \\
  \midrule   
  Diversity & Variety & Examines the breadth of perspectives or options presented by the idea, emphasizing its diversity. & \cite{chen2024asknaturenet_G10}, \cite{chung2022talebrush_W3} \\
  \midrule
  Functionality and Usability & Usability & Examines whether the idea is intuitive and practical, facilitating ease of adoption and application. & \cite{wang2023task_D8}, \cite{chung2022talebrush_W3} \\
  \cmidrule{2-4} 
  & Feasibility & Determines whether the idea is realistically implementable in practical scenarios. & \cite{wang2023task_D8}, \cite{pu2024ideasynth_S1} \\
  \cmidrule{2-4}  
  & Efficiency & Assesses the idea’s potential to achieve objectives effectively within resource constraints. & \cite{chung2022talebrush_W3} \\
  \cmidrule{2-4} 
  & Functionality & Evaluates the idea’s capacity to address specific functional needs or solve targeted problems. & \cite{wang2023task_D8} \\
  \midrule
  Impact and Inspiration & Impact & Measures the potential of the idea to effect significant change or address major challenges. & \cite{pu2024ideasynth_S1}, \cite{banker2024machine_S4} \\ 
  \cmidrule{2-4}
  & Insightfulness & Assesses whether the idea provides deep, unique insights that enhance understanding of a problem. & \cite{shaer2024ai_G1} \\
  \cmidrule{2-4}  
  & Inspiration & Evaluates the idea’s ability to stimulate further creative thinking or ideation. & \cite{chung2022talebrush_W3} \\
  \cmidrule{2-4} 
  & Meaningfulness & Considers whether the idea is contextually significant and capable of evoking value or resonance. & \cite{bhavya2023cam_G16} \\
  \midrule 
  Logical Coherence & Plausibility & Determines whether the idea is logical, realistic, and consistent with common sense or knowledge. & \cite{banker2024machine_S4} \\
  \midrule  
  Clarity and Specificity & Clarity & Assesses whether the idea is clearly articulated and unambiguous, enhancing its comprehensibility. & \cite{banker2024machine_S4} \\
  \cmidrule{2-4}  
  & Specificity & Evaluates the level of detail and precision in the idea, ensuring actionable outcomes. & \cite{pu2024ideasynth_S1} \\
  \midrule 

  \bottomrule
  \end{tabular}
  }
\end{table}

Functionality and usability-related criteria, such as usability, feasibility, efficiency, and functionality, were used in a total of 6 studies, with usability and feasibility being the most common (2 studies each). Impact and inspiration-related criteria, including impact, insightfulness, inspiration, and meaningfulness, were used in 5 studies, with impact being the most frequent (2 studies).

Other evaluation criteria, such as plausibility, clarity, and specificity, were each used in one study, highlighting the diverse range of factors considered when assessing the quality and effectiveness of ideas generated with the assistance of LLM-based systems.

\newpage
\section{Discussion}
This section synthesizes the primary findings of our review and outlines promising directions for future research in LLM-assisted ideation. We first provide an overview of our key results, then discuss detailed observations from our analysis of the ideation framework, interaction design, and user study, and finally describe several research gaps.

\subsection{Summary}
In this section we present our findings in terms of the ideation framework, interaction design, and user study, but first we present three general conclusions.

\subsubsection{\textbf{Overview}}

Our review of 61 studies reveals several overarching conclusions regarding LLM-assisted ideation:

\begin{itemize}
    \item [1.] \textbf{Rapid Growth.} LLM-assisted ideation is an emergent field, with publication counts rising from 5 studies in 2022 to 37 in 2024. The most influential publication venues include CHI, UIST, and IUI.
    \item [2.] \textbf{Diverse Application Domains.} The research covers five principal application areas: design, writing, multimedia creation, scientific research, and education.
    \item [3.] \textbf{Integration with Ideation Methods.} Beyond specific applications, studies have integrated LLMs with various ideation techniques—such as structured thinking, decomposition/composition, analogical thinking, and collaborative ideation—to enhance the ideation processes.    
\end{itemize}

\subsubsection{\textbf{Ideation Framework}}

Our analysis of the ideation framework reveals significant variability in LLM adoption across different stages:

\begin{itemize}
    \item [1.] \textbf{Scope Definition}: Recognized as a critical starting point (reported in 97\% of studies), yet LLM involvement is minimal (5\%).
    \item [2.] \textbf{Foundational Materials Collection and Structuring}: Addressed in 39\% and 16\% of studies, respectively, with high LLM adoption rates (67\% and 90\%). Notably, no studies explored visualizing structured materials.
    \item [3.] \textbf{Idea Generation}: Present in all studies (100\%), with 75\% relying on LLMs for solely direct idea generation. However, relatively few studies employed LLMs to stimulate human creativity but instead provided immediately usable ideas.
    \item [4.] \textbf{Idea Refinement}: Featured in 64\% of studies, with 87\% demonstrating that LLMs can enhance creative refinement.
    \item [5.] \textbf{Idea Evaluation \& Multi-idea Evaluation and Selection}: Idea evaluation was addressed in 75\% of studies, yet only 46\% utilized LLMs in this stage. Few studies incorporated evaluation tools (e.g., PPCO). Multi-idea evaluation and selection (present in 46\% of studies) saw limited LLM support (14\%). Few studies considered structuring multiple ideas, especially in team ideation scenarios.
    \item [6.] \textbf{Iterative Ideation}: Engaged in only 46\% of the studies, primarily focusing on core stages (idea generation and refinement) but also impacting earlier stages (scope definition and foundational material collection).
\end{itemize}

These findings highlight an imbalance in current LLM-assisted ideation research, with a disproportionate emphasis on divergent phase. Preparation and convergent phases would benefit from further investigation and work.

\subsubsection{\textbf{Interaction Design}}

Our review indicates that LLM-assisted ideation systems, which can be broadly classified into three categories: tools (56\%), frameworks (11\%), and methods (33\%), predominantly support individual creativity (85\% of studies), with limited attention to group-based activities (e.g., research landscaping meetings). Key observations include:

\begin{itemize}
    \item [1.] \textbf{Modality of Interaction}: Text-based interaction dominates (98\%), suggesting untapped potential for diverse modalities and multimodal interactions.
    \item [2.] \textbf{Interface Element}: Common interface components include canvas-based tools (44\%), text editors (32\%), and panels (32\%).
    \item [3.] \textbf{Idea Management}: 17 tool interfaces (50\%) develop idea functional components, including adding, deleting, and editing ideas, considering users’ ability to manage ideas. 
    \item [4.] \textbf{Human–LLM Interaction}: Open interaction is prevalent in 46\% of studies, with triggered functions (43\%) and user-defined prompts (26\%) frequently used. Although approximately half of the tools incorporate creative management capabilities, few provide intelligent assessments of creative quality or scope alignment.
\end{itemize}

These trends underscore the opportunity to expand beyond text-centric, individual-oriented designs toward richer, multimodal, and collaborative interfaces, with a greater emphasis on designing system idea management modules and human-computer interaction patterns tailored to specific tasks.

\subsubsection{\textbf{User Study}}

Approximately 70\% of the reviewed papers incorporated user studies with diverse experimental designs and evaluation protocols:

\begin{itemize}
    \item [1.] \textbf{User Study Design}: 
        \item Single-condition evaluations were used in 42\% of studies, while 40\% compared two conditions.
        \item Within-subjects designs (77\%) were more common than between-subjects designs (19\%).
    \item [2.] \textbf{Analytical Methods}: 
        \item Qualitative approaches were employed in 79\% of studies, predominantly via semi-structured interviews and inductive analysis.
        \item Quantitative techniques (63\%) typically involved Likert scale ratings, with both descriptive (70\%) and inferential (70\%) statistics applied.
    \item [3.] \textbf{Assessment Criteria}: Evaluation of ideation quality         frequently focused on novelty, functionality, inspirational impact,         relevance, and diversity.
\end{itemize}

These evaluation protocols provide a useful reference for researchers seeking to assess the efficacy and impact of LLM-assisted ideation systems. 

\newpage
\subsection{Research Gaps}
Based on our analysis, we propose several avenues for future research in LLM-assisted ideation, structured according to the dimensions of the ideation framework, interaction design, and methodological approaches.

\subsubsection{\textbf{Ideation Framework}}

We identify ten key recommendations:

\begin{itemize}
    \item [1.] \textbf{Articulation of Creative Goals}: Develop techniques enabling LLMs to help users articulate creative goals more clearly and comprehensively.
    \item [2.] \textbf{Scope-Driven Evaluation}: Explore LLM-based methods to assess idea relevance to predefined scopes, enhancing evaluation efficiency and effectiveness.
    \item [3.] \textbf{Visualization Techniques}: Explore visualization of structured foundational materials to enhance rapid comprehension and at-a-glaze analysis to facilitate and enrich group discussions.
    \item [4.] \textbf{Free-Riding Detection}: Explore methods to detect both conscious and unconscious free-riding during creative ideation to improve LLMs assistance or stimulation stages.
    \item [5.] \textbf{Adjustable Assistance}: Develop strategies for dynamically adjusting the degree of LLM assistance and stimulation during the generative process across idea generation, refinement, and evaluation stages.
    \item [6.] \textbf{Integration of Professional Tools}: Incorporate established evaluation frameworks (e.g., PPCO, SWOT) into LLM-assisted evaluation, particularly for group ideation.
    \item [7.] \textbf{Trust and Adoption}: Examine underlying trust issues in the limited LLM adoption in the idea evaluation and selection stages to improve people’s trust in LLM capabilities.
    \item [8.] \textbf{Balancing Judgment}: Investigate the optimal balance between human judgment and LLM-generated insights in evaluation and selection, and ways to provide adjustment or incremental LLM support.
    \item [9.] \textbf{Multi-idea Structuring}: Explore methods for structuring multiple ideas in LLM-assisted multi-idea evaluation and selection, this is quite valuable due to the volume of refinement or evaluation items generated and time constraints in group ideation.
    \item [10.] \textbf{Iterative Support Mechanisms}: Explore support mechanisms that facilitate iterative ideation processes tailored to specific activities or domains.
\end{itemize}

\subsubsection{\textbf{Interaction Design}}

We have identified three primary areas of development with regard to how interactivity is currently being embedded in such tools:

\begin{itemize}
    \item [1.] \textbf{Group Ideation Support}: Investigate methods for LLMs to address the unique challenges of group ideation, including fostering effective group collaboration, ensuring transparent and equal participation, and facilitating consensus-building.
    \item [2.] \textbf{Multimodal Interactions}: Evaluate the relationship between creative tasks and optimal interaction modes, by integrating multimodal interactions to expand application areas and improve user experience or exploring innovative interface components and combinations, particularly those based on canvas interfaces.
    \item [3.] \textbf{Impact on User Experience}: Examine the effects of various human–LLM interaction formats on user experience, creativity, and cognitive load as LLM technologies become increasingly pervasive.
\end{itemize}

\subsubsection{\textbf{Ideation Methodology}}

Additional research is warranted to advance the foundations of LLM-assisted ideation:

\begin{itemize}
    \item [1.] \textbf{Framework Development}: Creating theoretical frameworks to guide the selection and integration of LLM capabilities into established ideation methods will be essential. Existing ideation techniques, such as brainstorming, lateral thinking, and morphological analysis, could be enhanced by incorporating LLMs at appropriate stages. Frameworks are needed to systematically identify opportunities for LLM assistance and optimize the human-LLM collaboration.
    \item [2.] \textbf{Taxonomy and Mapping}: Developing a taxonomy of ideation problems and mapping them to ideation solution methods, which include iterative ideation frameworks, human-computer interaction paradigms, and LLM-assisted techniques will also be critical. This could involve categorizing ideation tasks based on factors such as domain, complexity, level of abstraction, and required background knowledge or dataset. This taxonomy could guide the design and development of ideation processes and their iterative workflows for specific ideation problems, as well as inform the creation of unique human-computer interaction modes that are most suitable for addressing specific problem. Establishing this mapping helps to rapidly construct ideation solutions tailored to specific problems, determining when and how LLMs should be employed in the ideation process.
\end{itemize}

\newpage
\section{Conclusions}

Recent advancements in large language model (LLM) technologies have catalyzed significant interest in their application to ideation processes across various creative domains. In this paper, we synthesized 61 studies examining LLM-assisted ideation, providing an overview of publication trends, application domains, and methodologies.

We proposed an iterative ideation framework that comprised three phases: preparation, divergent, and convergent, and identified seven key stages: scope definition, foundational materials collection, foundational materials structuring, idea generation, idea refinement, idea evaluation, and multi-idea evaluation and selection. Our analysis indicates that LLMs are predominantly leveraged for idea generation and refinement, whereas their use in evaluating and selecting multiple ideas remains limited.

Furthermore, our review of interaction design and user studies reveals a prevailing focus on individual ideation activities and text-based interactions, with emerging trends toward multimedia integration. However, in group ideation, tools and interaction modalities targeting both synchronous and asynchronous collaboration are much scarcer.

The insights derived from this review underscore the substantial potential of LLM-assisted ideation while highlighting several research gaps. Future work could concentrate on developing more effective and tailored methods for harnessing LLM capabilities, particularly in convergent phases and collaborative settings, in order to enhance creative problem-solving and drive innovation across diverse fields.

\section{Open Data, Tools, and Collaborative Future Directions}

We are pleased to announce that the complete analysis, including all referenced publications and associated data, is now available online\footnote[3]{Resources can be accessed at \url{https://doi.org/10.6084/m9.figshare.28440182}}. This interface is designed to enable researchers to explore our findings comprehensively, thereby fostering new research initiatives and extending the collective knowledge in the field of LLM-assisted ideation.

We envision that this openly accessible repository will serve not only as a reference but also as a collaborative platform. Researchers are encouraged to fork the provided information, refine the methodologies, and build upon our work to further enhance creative problem-solving through improved human–LLM collaboration. By embracing iterative contributions and cross-disciplinary perspectives, we anticipate that the field will continue to evolve dynamically, driving innovation and enabling more sophisticated applications in both individual and group ideation processes.

In the spirit of collaborative advancement, we hope that our shared resources inspire rigorous inquiry and innovative extensions, ultimately contributing to a deeper understanding and broader application of LLM-assisted ideation in diverse domains.

\section*{Disclosure statement}
No potential conflict of interest was reported by the author(s).

\newpage
\printbibliography

\end{document}